# Conceptual design study for heat exhaust management in the ARC fusion pilot plant

A.Q. Kuang[1], N.M. Cao[1], A.J. Creely[1], C.A. Dennett[2], J. Hecla[2], B. LaBombard[1], R.A. Tinguely[1], E.A. Tolman[1], H. Hoffman[1], M. Major[1], J. Ruiz Ruiz[1], D. Brunner[1], P. Grover[3], C. Laughman[3], B.N. Sorbom[1], D.G. Whyte[1]

[1]MIT Plasma Science and Fusion Center, Cambridge MA 02139 USA
[2]MIT Nuclear Science and Engineering Department, Cambridge MA 02139 USA
[3]Mitsubishi Electric Research Laboratories, Cambridge MA 02139 USA

# Conceptual design study for heat exhaust management in the ARC fusion pilot plant

A.Q. Kuang[1], N.M. Cao[1], A.J. Creely[1], C.A. Dennett[2], J. Hecla[2], B. LaBombard[1],
R.A. Tinguely[1], E.A. Tolman[1], H. Hoffman[1], M. Major[1], J. Ruiz Ruiz[1],
D. Brunner[1], P. Grover[3], C. Laughman[3], B.N. Sorbom[1], D.G. Whyte[1]

[1]MIT Plasma Science and Fusion Center, Cambridge MA 02139 USA
[2]MIT Nuclear Science and Engineering Department, Cambridge MA 02139 USA
[3]Mitsubishi Electric Research Laboratories, Cambridge MA 02139 USA

The ARC pilot plant conceptual design study has been extended beyond its initial scope [B. N. Sorbom *et al.*, FED **100** (2015) 378] to explore options for managing ~525 MW of fusion power generated in a compact, high field ($B_0$ = 9.2 T) tokamak that is approximately the size of JET ($R_0$ = 3.3 m). Taking advantage of ARC's novel design – demountable high temperature superconductor toroidal field (TF) magnets, poloidal magnetic field coils located inside the TF, and vacuum vessel (VV) immersed in molten salt FLiBe blanket – this follow-on study has identified innovative and potentially robust power exhaust management solutions. The superconducting poloidal field coil set has been reconfigured to produce double-null plasma equilibria with a long-leg X-point target divertor geometry. This design choice is motivated by recent modeling which indicates that such configurations enhance power handling and may attain a passively-stable detachment front that stays in the divertor leg over a wide power exhaust window. A modified VV accommodates the divertor legs while retaining the original core plasma volume and TF magnet size. The molten salt FLiBe blanket adequately shields all superconductors, functions as an efficient tritium breeder, and, with augmented forced flow loops, serves as an effective single-phase, low-pressure coolant for the divertor, VV, and breeding blanket. Advanced neutron transport calculations (MCNP) indicate a tritium breeding ratio of ~1.08. The neutron damage rate (DPA/year) of the remote divertor targets is ~3-30 times lower than that of the first wall. The entire VV (including divertor and first wall) can tolerate high damage rates since the demountable TF magnets allow the VV to be replaced every 1-2 years as a single unit, employing a vertical maintenance scheme. A tungsten swirl tube FLiBe coolant channel design, similar in geometry to that used by ITER, is considered for the divertor heat removal and shown capable of exhausting divertor heat flux levels of up to 12 MW/m$^2$. Several novel, neutron tolerant diagnostics are explored for sensing power exhaust and for providing feedback control of divertor conditions over long time scales. These include measurement of Cherenkov radiation emitted in FLiBe to infer DT fusion reaction rate, measurement of divertor detachment front locations in the divertor legs with microwave interferometry, and monitoring "hotspots" on the divertor chamber walls via IR imaging through the FLiBe blanket.

# Contents

# 1. Introduction

A recent study [1] provided a conceptual design for a compact fusion pilot plant and fusion nuclear science facility based on the exploitation of newly-available high-temperature superconductor (HTS) tapes for high-field magnets. The ARC (Affordable Robust Compact) design has several attractive features including:

1. plasma energy gain Q > 10 at compact size (major radius, R ~ 3.3 m) due to high magnetic fields (peak B on coil ~ 23 T, magnetic field on the plasma axis, $B_0$ ~ 9.2 T) enabled by the use of REBCO (rare-earth barium copper oxide) HTS,
2. demountable toroidal field coils for vertical, modular replacement of interior components – a feature permitted by the improved material thermal and cooling properties at the higher operating temperature range (~20-30 K) enabled by REBCO,
3. a thin, replaceable, actively-cooled modular vacuum vessel (VV) with a conformal, close fit to the plasma poloidal shape, and
4. a liquid immersion blanket of FLiBe molten salt – which completely surrounds the VV; this high-temperature, single-phase, low-pressure fluid serves as (1) an effective medium for neutron moderation, heat removal, shielding and capture for efficient tritium breeding and (2) a large thermal reservoir to be directed to cool first-wall and VV components.

A key design challenge not addressed in the original ARC study was plasma power exhaust. This concern naturally arises due to the high global power density; as in the ITER design [2], ~500 MW of fusion power are produced, but in ~1/8$^{th}$ the volume of ITER. This high global power density is an attractive feature for a pilot fusion device – and likely necessary for extrapolating to commercial fusion power devices. However, a concern is that ARC may not have the ability to adequately exhaust ~150 MW of fusion alpha heating plus external radio-frequency (RF) heat deposited into its compact core. As discussed in more detail below, this concern is not unique to ARC; it is generic to all fusion reactor designs that must attain a certain neutron flux density (e.g., Fusion Nuclear Science Facility, FNSF) or maximize core fusion power density for economic considerations. In this regard, ARC's high magnetic field enables neutron wall loading that is similar to these large reactor designs but with smaller volume and reduced total power [3]. The net effect is that heat flux densities entering into ARC's divertor are projected to be no worse than the larger reactor designs. In any case, new robust power exhaust management solutions are needed.

The original ARC conceptual design study did not include an assessment of potential divertor solutions. However, some specialized features of the boundary/divertor plasma were examined [1]. The use of high-field side radio frequency (HFS-RF) launchers was explored, which had been shown previously to be attractive for efficient current drive and current profile control [4]. Recent studies of HFS-RF have further highlighted potential advantages: decreased launcher neutron damage and improved tritium breeding [5], efficient lower-hybrid current drive and ICRF heating/current drive due to improved accessibility [6], and enhanced impurity screening of HFS impurity sources [7,8]. The original ARC design study also performed neutron transport calculations to assess neutron shielding and tritium breeding [1]. The potential impact of a tungsten divertor target on neutronics and tritium breeding was assessed by attaching tungsten panels to the plasma-facing side of the VV where the lower divertor would be located; yet this was not meant to be a realistic divertor target plate geometry. Finally, the ability to actively cool a double-walled VV with FLiBe was assessed with thermal hydraulic calculations and simulations. This coolant channel design was found capable of handling both volumetric heating of the VV from neutrons and surface heating of the VV wall from core plasma photon radiation.

With regard to developing a plasma power exhaust management system, the original design study clearly identified ARC's unique design advantages:

1. the attractiveness of FLiBe as a coolant due to its very large temperature window in the liquid state, ~700-1700 K, and volumetric heat capacity, which allows for single-phase, low-pressure heat transfer,
2. the geometric simplification and advantage of having an extremely large fluid heat sink, i.e. the liquid blanket, immediately adjacent to the solid components, the VV and plasma-facing components (PFCs), which require active cooling, and
3. the low electrical conductivity of FLiBe, and its moderate viscosity, which greatly reduce magneto hydrodynamics (MHD) effects in fluid flow and enable sufficient fluid flow rates at modest pressure drop and pumping power.

The goal of this follow-on study is to explore and exploit these features for heat exhaust management with a conceptual design that includes an advanced divertor, while also retaining the essential features of the original ARC design: overall plasma geometry (major radius, minor radius, elongation), double-null divertor magnetic topology, demountable toroidal field (TF) coils, FLiBe liquid immersion blanket, and 525 MW DT fusion power. It is important to note that the goal of this study is not to develop a complete, self-consistent design for a power management system but to explore novel approaches made possible by ARC's unique design and to assess them from basic engineering considerations. It is also important to note that this study only explores potential options for *steady state* power management; start-up and shut down phases of the reactor and the impact of disruptions are not considered, for example.

Research performed over the past several years has clearly identified the challenges facing steady state heat exhaust management for tokamak power reactors, and they are daunting – in particular the requirement of maintaining a dissipative, cold divertor plasma in the presence of a narrow heat flux channel that is projected to be less than 1 mm wide for the poloidal magnetic fields of a power reactor [9,10]. At the same time, promising results have come from theoretical [11-13] and experimental [14-16] explorations of advanced divertor geometries, which indicate that such configurations may be able to meet this challenge. Advanced divertor geometries generally feature extended volumes for the divertor, additional poloidal field nulls, and shaping control beyond what is used for a standard vertical target divertor, as employed by ITER. However, Lackner and Zohm [17] have cautioned that currents in the poloidal field (PF) coils needed to create some of these configurations could be a major technological challenge – if the PF coils are placed far from the plasma to accommodate neutron shielding and/or outside the TF coils due to assembly constraints. TF coils produced with low-temperature superconductors (e.g. $Nb_3$-Sn) cannot be segmented and therefore, as a practical consideration, such superconducting PF coils must be placed outside the TF. Because ARC has demountable TF coils, superconducting PF coils can be placed *inside* the TF, much closer to the plasma allowing for much lower total coil currents and forces.

The placement of PF coils inside the TF is an approach taken by many tokamaks, such as DIII-D [18], Alcator C-Mod [19] and TCV [20]; these physics experiments utilize this feature to enhanced plasma shaping capabilities. A principal difference is that ARC also has a 1-meter-thick FLiBe blanket surrounding its plasma core for neutron shielding. We find that a long-legged, advanced divertor can be implemented in ARC simply by carving out an appropriate space in the FLiBe blanket. Thus, the implementation of an advanced, long-legged divertor in a power-producing DT fusion reactor does not necessarily require a decrease in core plasma volume and/or increase in TF magnet size.

The present study explores the implementation of a X-point Target (XPT) divertor in ARC, similar to the XPT configuration proposed for the ADX divertor test tokamak [13]. In this configuration, an additional (secondary) poloidal magnetic field null is formed at the end of an extended outer divertor leg in both the lower and upper divertors. Recent modeling by Umansky *et al*. [21] has shown that this configuration may attain a stable, highly dissipative (i.e., fully detached) divertor, accommodating upstream power densities that are ~10 times higher than for standard vertical target divertors. Plasma fluid modeling is presently underway for the divertor configuration specifically chosen in this follow-on ARC study and will be published in a separate paper [22]. Initial results indicate that a stable, fully detached divertor plasma state may be achieved even at the maximum levels of divertor power exhaust anticipated (~105 MW) while only requiring minimal levels of impurity seeding (e.g., ~0.5% neon, fixed fraction). While the present design study focused on implementing an XPT for ARC, this design exercise clearly indicates that other advanced divertor ideas that utilize a long legged geometry (e.g. super-X divertor [23]) may be implemented as well.

The organization of this paper is as follows. Section 2 reviews power exhaust requirements for ARC relative to other reactor designs, taking into consideration the empirically observed poloidal magnetic field scaling of scrape off layer power widths and the economic needs of attaining a certain neutron power loading and/or core fusion power density. Contrary to initial expectations, scaling considerations reveal that heat flux densities entering into the divertor of ARC will be no worse than those anticipated for other reactor designs such as ARIES [24], ACT1 and ACT2 [25], and European DEMO [26], which in general are much larger devices operating at moderate magnetic fields.

Section 3 examines requirements for the placement of PF coils to achieve the XPT divertor configuration for an ARC magnetic equilibrium. Superconducting poloidal field coil locations are specified with consideration of neutron shielding, current densities in the HTS, forces on the coils and the attainment of the XPT divertor magnetic geometry. The resultant modification to the VV shape impacts other aspects of heat exhaust management, such as requirements for plasma position and shaping control, and consideration of neutron damage to high heat flux divertor target components.

Neutron transport in the modified VV and blanket configuration is addressed in Section 4, with a focus on obtaining the required shielding for the inner PF coils and maintaining the tritium breeding ratio (TBR). This study also examines the neutron flux at primary divertor surfaces – in particular its softened energy spectrum – which is a benefit of geometrically separating and shielding these surfaces from line-of-sight view to the core plasma neutron source.

Taking advantage of the favorable features of FLiBe listed above, combined with new ideas for cooling channel geometries afforded by advanced manufacturing, Section 5 explores novel design approaches for an integrated heat exhaust management system. Schemes to implement forced flow cooling of vacuum vessel and divertor components are developed, capable of handling ARC's full thermal output while minimizing pumping power. The divertor cooling system borrows embedded swirl tubes from the ITER design [27], but with FLiBe coolant. The design is shown capable of removing surface heat fluxes of up to 12 MW/m$^2$. Although the designs presented here are only conceptual, they show that the required levels of local and global heat transfer can be obtained in ARC with acceptable coolant pumping power (< 1% of fusion power) – a necessary requirement for an economical fusion power plant.

Finally, Section 6 explores the issues of plasma detachment control and diagnostics that are enabled by unique aspects of the ARC reactor design. One of the key advantages of the XPT divertor, the large detachment power window [21], is explored with respect to steady state reactor operation. In addition, the long leg divertor geometry lends itself to a feedback control system of the detachment front location using microwave interferometry as a detector. Lastly, the optical transparency of the FLiBe blanket makes possible diagnostics that 'look through' the blanket: (1) thermal imaging of the external surface of the VV, (2) monitoring of the Cherenkov radiation produced in FLiBe to deduce the fusion reaction rate.

As in any conceptual design study, the overall goal is to identify not only what might be possible but also what needs further study. Key areas needing further investigation include: quantifying the power handling performance of the X-point target divertor; assessing and developing advanced manufacturing techniques, such as additive manufacturing, to construct the designs considered; employing advanced computational fluid dynamics for refined heat transfer analysis; quantifying the effects of neutron damage to HTS superconductors; and fully characterizing the optical properties of FLiBe, particularly in the radiation environment of a DT fusion reactor. These are discussed in Section 7.

## 2. Assessment of the divertor heat flux challenge for ARC compared to lower field reactor designs

A critical challenge for any fusion power plant design is to implement a robust divertor system that can handle extreme power densities and also suppress damage to target surfaces that would otherwise arise from energetic particle sputtering and helium implantation. Lacking exotic solutions, such as liquid metal and/or vapor targets, this calls for the attainment of a stable, completely detached divertor plasma state. Based on the current understanding of plasma transport in the scrape-off layer (SOL), key parameters that determine the plasma temperature and density in the divertor – and thus the access window for detachment – are the midplane plasma pressure and peak parallel heat flux. The plasma pressure at the midplane is more or less fixed by core confinement requirements for optimal fusion operation and is ubiquitous for all reactor fusion plasmas. The peak parallel heat flux is determined by cross-field energy transport mechanisms in the SOL, which are not yet quantifiable from first principles models. Recent measurements from multiple experiments provide some guidance. They have led to the 'Eich scaling' empirical relationship [9], which indicates that the upstream midplane power exhaust width scales inversely with poloidal magnetic field strength and is insensitive to reactor size. Based on these observations, one might conclude that a high-field compact reactor like ARC, (operating with a 1.5 T poloidal magnetic field with a ~0.4 mm boundary heat flux width) must operate at higher parallel heat fluxes in comparison to low-field large reactor designs such as ARIES [24], ACT1 and ACT2 [25], and European DEMO designs [26], which are already projected to have unmitigated parallel heat flux power densities entering their divertors exceeding 40 GW/m$^2$. This raises a key question for the ARC design: Would the attainment of a fully detached state be inherently more difficult? In this regard, one must also consider the total power that must be exhausted into the SOL in these designs.

ARC features relatively high volumetric power density ~3.7 MW/m$^3$ from 525 MW of fusion power produced in a 137 m$^3$ plasma volume. This is accomplished below the no-wall beta limit and at high safety factor ($q_{95}$~7) by taking advantage of higher field with the REBCO superconductors and fusion power density scaling as ~B$^4$. Yet, in comparison to the reactor designs listed above, ARC has a ~2-3 times higher surface area to volume ratio due its small size (R), which keeps its global areal power density at ~2.5 MW/m$^2$, similar to the larger reactors. This situation is attractive from an economic point of view since materials cost is approximately proportional to volume, which means ARC is a lower entry point cost for a pilot plant since it is ~5-20 times smaller in volume than the other designs.

Based on the Eich scaling [9], the parallel heat flux entering into a divertor is proportional to PB/R [28] where P is the total heating power (~$P_{fusion}$/5 at high gain), B is the toroidal magnetic field, and R is the major radius. At first glance this scaling appears highly unfavorable for high B, small R devices. However, one must also consider that P itself is constrained based on the device's mission. In a FNSF, P/S ~ P/R$^2$, is fixed by design in order to provide sufficient neutron flux density to test components. In a power plant P/V ~ P/R$^3$, is fixed in order to meet power/cost economic targets. Thus, at high fusion power gain, the parallel heat flux in these devices scales as either $q_{//}$ ~ BR (FNSF) or $q_{//}$ ~BR$^2$ (power plant). The Eich scaling is therefore not necessarily punitive for heat exhaust in small R, high B designs, when the global power density requirements are otherwise held fixed.

This somewhat counter-intuitive scaling, i.e. *that small high field devices will improve the heat exhaust situation*, essentially stems from the lack of any explicit size dependence in the Eich scaling. A more detailed study by Reinke [29] shows that access to a dissipative divertor is enabled by high B, in part due to the access to higher core densities due to the empirical density limit, $n_{Greenwald}$ ~ I/R$^2$ at fixed aspect ratio, which disallows high density in large devices. In the case of ARC, which is best described as an FNSF, it turns out the $q_{//}$ ~ BR scaling is approximately followed because the total fusion power P is smaller than the large reactors listed above yet the parallel heat flux is similar (12 to 40 GW/m$^2$). So, to a first approximation, power exhaust in ARC is neither more nor less difficult than in much larger devices with lower field; yet the motivation to increase B and decrease R is clear from a costing view. We therefore treat the power exhaust challenge in ARC as being generically similar to other reactor designs and, in any case, seek to implement the most robust divertor system available.

## 3. Magnetic equilibrium with X-point target divertor

Recognizing the need to employ a very robust, high power density handling divertor solution for ARC, a long-leg XPT divertor configuration, combined with operation in balanced double null magnetic equilibria, was deemed highly desirable [28]. Recent analyses found this configuration to be extremely promising [21], attaining a factor of ~10 increase in power density compared to conventional divertors operating at the same upstream plasma conditions. In addition, the modeling shows that a stable, fully detached divertor state may be maintained over a large power window while avoiding the formation of 'X-point MARFEs' [30], which degrades core plasma performance. For conventional reactor designs, it is difficult to implement the PF coil set needed for such advanced magnetic divertor topologies. When PF coils are located outside TF coils, very large coil currents may be required [17]. Located inside the TF coils, the PF coils can become difficult to shield from neutron heating and damage. However, the situation is very different in ARC. Its demountable TF magnet combined with its highly conformal, immersion FLiBe neutron blanket might allow the implementation of advanced divertor solutions.

### 3.1 Design goals

This design effort sought to attain the following goals for the magnetic equilibrium, divertor geometry and PF coil set:

1. Identify a PF coil set that will reproduce the core plasma magnetic equilibrium and shaping of the original ARC specifications (current, elongation, triangularity, major radius) while producing an XPT divertor configuration in a double-null geometry.
2. Achieve this within the envelope of the TF magnet, consistent with the original ARC specifications, i.e., *do not increase the size of the TF magnets*.
3. Locate superconducting PF coils outside the FLiBe tank, with adequate shielding for neutrons (i.e., approx. ~1 meter of shielding line-of-sight to core) but inside the HTS TF magnet envelop, taking advantage of its demountability.
4. Use high temperature superconductors for PF coils with appropriate cross-sections so as to attain acceptable limits of current density, considering the magnetic fields at the coils and published data on REBCO high temperature super conductor performance [31].
5. Fully utilize ARC's demountable TF coils, maintaining the ability to lift the integral vacuum vessel, divertor and cooling system assembly from the FLiBe shielding tank as a single unit for maintenance and replacement, without disturbing the superconducting PF coil set.
6. Locate magnetic sensors in regions with adequate shielding against neutrons yet with sufficient time response and sensitivity to changes in magnetic equilibrium.
7. If non-superconducting coils are needed inside the neutron shield, identify potential approaches that can meet the requirements of thermal loading and power dissipation.
8. Examine the overall forces on the PF coil set and identify a means to accommodate them.

In addition, the design sought to investigate and to potentially exploit the low electrical conductivity properties of the FLiBe blanket in order to:

1. potentially provide low-voltage electrical isolation for non-superconducting trim coils located within the FLiBe tank, and
2. potentially allow magnetic flux to penetrate through the blanket on short time scales. This could reduce the time response for poloidal flux sensors located on the outside of the shielding tank (making placement at this location an option) as well as reduce the penetration time and eddy current losses from fast poloidal flux swings from vertical stability control coils and divertor trim coils.

The output of this exercise is summarized in Figure 1. As described in detail in sections 3.2-3.5 below, we identified a magnetic equilibrium, divertor geometry, PF coil placement, and FLiBe tank geometry that could satisfy all of the design goals.

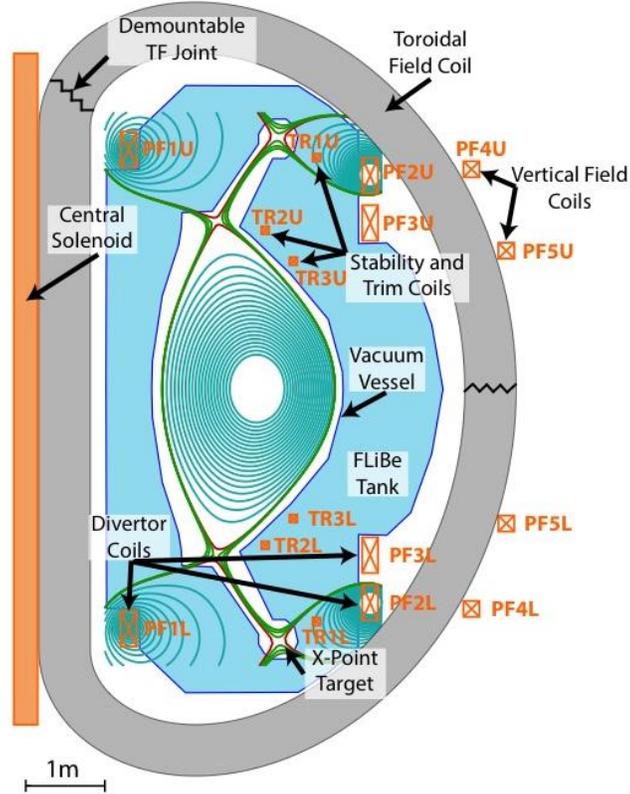

*Figure 1: Magnetic equilibrium, poloidal field (PF) and trim (TR) coil set, vacuum vessel and FLiBe tank geometry identified by the ARC-Divertor design study. The green poloidal flux contour lines are spaced 1mm apart at the outer midplane. The design accommodates a long-leg, XPT divertor in a balanced double null configuration with elongation and triangularity maintained very close to the original ARC design while fitting within the original toroidal field (TF) coil envelope. (For viewing clarity, the radial build [1] and TF joints are highly simplified).*

### 3.2 Magnetic equilibrium

Coil placements and magnetic equilibria were explored with the ACCOME MHD equilibrium code [32,33], using a customized GUI written in IDL to facilitate rapid scoping studies. This design was iterated in order to simultaneously meet all of the above requirements. First, a magnetic equilibrium was created that met the targeted requirements (balanced double null, divertor X-points, elongation, triangularity). The field at the coil locations was then calculated to determine the critical current density in the coils (discussed in Section 3.3) and thus the minimum coil size. This was then checked against the geometry of the FLiBe tank to ensure that the coils did not inhibit vertical removal of the vacuum vessel assembly. These results were then exported to perform a MCNP neutron transport calculation (discussed in Section 4) to ensure that the coils were sufficiently shielded to maintain a reasonable lifespan. The neutronics results were then used to guide the placement of the coils in the next iteration.

The final coil placement in this conceptual design is shown in Figure 1. It should be noted that this is a single point design; magnetic equilibria and PF coil currents for different values of central solenoid flux where not explored. Instead, the central solenoid geometry and flux were taken from the original ARC design. However, since ARC is envisioned to attain a fully non-inductive steady state condition (i.e. fixed central solenoid flux and PF currents), we consider this magnetic equilibrium as being appropriately representative for the design study. As specified in the original ARC paper, the design includes a plasma current of 8.0 MA, major radius of 3.3 m, elongation of 1.8, and triangularity of 0.375 [1]. The divertor coils create XPT divertors in the upper and lower divertor legs. The primary X-points are located at a major radius of 2.9 m while the secondary, divertor X-points are located at a major radius of 3.7 m. This provides a factor of 1.3 in total flux expansion (i.e. parallel flux bundle areal expansion) between primary and divertor X-points. In combination with neutral compression effects and plasma recycling on the divertor side walls [21], total flux expansion helps to stabilize thermal detachment fronts [34]. UEDGE simulations performed for an XPT divertor [21] based on the ADX device [28] find that this geometry can provide a wide operational power window for stable power exhaust, which is critical for heat flux management and control (discussed further in section 6).

The minimum plasma-wall gap was designed to be approximately 20 times the heat flux width (8 mm, mapped to the outer midplane). The present design takes advantage of significant poloidal flux expansion and a tilted plate geometry to reduce anticipated heat loads on the inner divertor target. Recent experiments in Alcator C-Mod found that approximately 10% of the power into the scrape-off layer goes to the inner divertor target plates in balanced double-null discharges (L, H, and I-mode) [35]. Thus, the simple inner divertor target geometry may be appropriate. Further optimization of first-wall surfaces should be explored, such as providing for tilting of inner and outer divertor target plates at strike point locations and more careful shaping at the entrance of the outer divertor chamber, but this exceeds the scope of the present study.

The magnetic equilibrium and divertor topology is achieved with divertor PF coil currents of 3.9, -4.4 and 5.2 MA, as indicated in Figure 2. These currents are achievable with new high performance superconductors. As shown in Table 1, all of these currents are well within the critical current density of REBCO HTS as they are sized in Figure 2 (to scale), even accounting for the high background magnetic field strengths. The PF coils are located just outside the FLiBe tank, and would require thermal insulation as described in [1]. The REBCO HTS critical current densities were calculated from the specifications found in [31]. As a baseline, the low end of the 'premium tape' was used ($J_e$ =275 A/mm$^2$ at 77 K and 0 T). Accounting for the different magnetic fields and temperature was done using the 'lift factors' included in the reference. 'Lift factor' is a magnetic field dependent parameter that quantifies the change in $J_e$ resulting from operation at temperatures lower than 77 K and fields greater than 0 T. For example at 20 K and 17 T a 'lift factor' of 1.27 was estimated resulting in $J_e$ =350 A/mm$^2$. This is a conservative estimate, since the tape performance has improved considerably since the publication of these specifications. Furthermore, the calculations utilize the total magnetic field strength at the coil location, from both toroidal and poloidal fields, applied along the most detrimental direction to tape performance (perpendicular to the tape direction). This again provides a conservative estimate of the coil critical current, as the actual angle of the field will slightly increase the available critical current density. In addition, these calculations assume an operating temperature of 20 K, as is the case with the toroidal field magnets [1]. Finally, the calculations assume that the coil will be 20% superconductor and 80% structure, which are conservative considering the relatively low hoop stresses applied to these coils, discussed further in Section 3.5. The coils have been sized such that they are large enough to stay below 65% of the critical current density (calculated using the assumptions stated), while still allowing vertical removal of the VV, greatly simplifying maintenance.

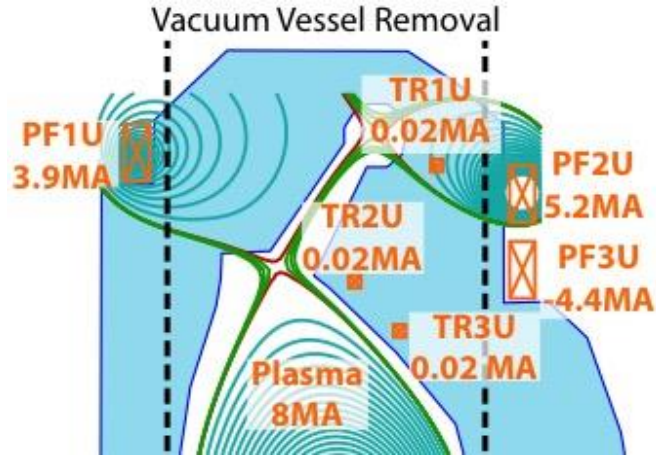

*Figure 2: Current and spacing requirements of HTS divertor coils, allowing vertical removal of the vacuum vessel. All coils operate at less than 6 MA, and are well within the critical current density of REBCO HTS. Internal copper trim (TR) coil locations and currents needed to vertical stability and divertor x-point control. Trim coil currents are below 20 kA, minimizing power consumption. Their close proximity to the plasma allows fast corrections to the magnetic geometry.*

In addition to equilibrium coils, the ARC divertor conceptual design also includes trim coils, which must respond to and correct for changes in the magnetic geometry on relatively fast time scales. The divertor X-point geometry presented here is sensitive to approximately ± 2% changes in the total plasma current, as well as approximately 5 cm vertical and radial displacements to the core plasma. Each of these events would cause the flux surface passing through the divertor X-point to change by 1 mm mapped to the outer midplane. The magnetic geometry must therefore be corrected on fast time scales in order to prevent an uneven distribution of heat loading to the divertor resulting in damage. Placing the trim coils inside of the FLiBe tank (Figure 2) reduces the current requirements of the coil and minimizes the amount of conducting material between the coils and the plasma, thus reducing the field propagation time to ~50 ms, as discussed in section 3.4. However, this placement also means that the coils must be designed to survive the high temperature and large neutron flux inside of the FLiBe tank. The trim coils require a maximum ~20 kA to correct the magnetic perturbations discussed above, so instantaneous power consumption is minimized (~25 kW per coil), although time-averaged power consumption is assumed to be negligible since these coils are activated temporarily to address equilibria shifts. Note that due to the narrow heat flux width, a vertical displacement in the plasma on the order of a millimeter can result in a loss of balanced double null and a significant increase in heat-flux to the dominant outer divertor or, even worse, to an inner divertor surface. Such displacements are a challenge to detect and correct reliably with magnetic sensors. Section 6 discusses the possibility of measuring the power sharing of the upper and lower divertors directly by sensing the detachment front positions in the outer divertor legs. This in turn can be used as a means to feedback control the plasma position and magnetic flux balance, using the parameters that matter most – divertor power loading.

All coil locations, sizes and currents are summarized in Table 1.

*Table 1: Poloidal field coil design specifications*

|  | Divertor coils | | | Vertical field coils | | Trim coils | | |
| --- | --- | --- | --- | --- | --- | --- | --- | --- |
|  | **PF1** | **PF2** | **PF3** | **PF4** | **PF5** | **TR1** | **TR2** | **TR3** |
| **Conductor** | REBCO | | | | | Cu | | |
| **Current (MA)** | 3.9 | 5.2 | -4.4 | -1.4 | -3.5 | +/- 0.02 | +/- 0.02 | +/- 0.02 |
| **R (m)** | 1.80 | 4.85 | 4.85 | 6.13 | 6.57 | 4.20 | 3.59 | 3.98 |
| **Z (m)** | 3.03 | 2.70 | 2.10 | 2.78 | 1.70 | 2.85 | 1.97 | 1.70 |
| **Width (m) (radius (m) for trim coils)** | 0.20 | 0.20 | 0.20 | 0.20 | 0.20 | 0.11 | 0.11 | 0.11 |
| **Height (m)** | 0.45 | 0.45 | 0.45 | 0.20 | 0.20 | NA | NA | NA |
| **$B_r$ (T)** | 0.77 | 1.96 | 1.30 | 0.83 | -0.01 | NA | NA | NA |
| **$B_z$ (T)** | 0.12 | -0.49 | -1.26 | -0.71 | -0.10 | NA | NA | NA |
| **$B_{toroidal}$ (T)** | 17.0 | 6.29 | 6.29 | 4.98 | 4.64 | 7.27 | 8.50 | 7.67 |
| **Current density (A/mm$^2$)** | 43.3 | 57.8 | 48.9 | 35.0 | 87.5 | 0.5 | 0.5 | 0.5 |
| **Critical coil current density, 20% REBCO tape [31] (A/mm$^2$)** | 70.0 | 95.4 | 95.8 | 114.4 | 122.4 | NA | NA | NA |
| **Resistive power loss at full current (kW)** | NA | NA | NA | NA | NA | 20 | 20 | 20 |

### 3.3 Internal trim coils

Low-current trim coils are employed in the ARC design to respond to and correct for plasma shape changes on fast time scales for vertical stability and divertor X-point control (Figure 2). The location of these coils inside the FLiBe tank, which operates at ~900 K, and close to the plasma in a region of high neutron fluence, precludes the use of superconductors. For this reason, we have considered a copper conductor design. These coils are considered replaceable along with the VV. We seek to determine if a design window exists for a copper coil that can withstand the high corrosion and the high temperatures of FLiBe in this environment – exacerbated by high electrical resistivity leading to ohmic losses, and neutron induced volumetric heating. For this purpose, we consider a heat transport analysis using a highly simplified copper conductor clad in Inconel, to determine whether the copper will melt. We consider a solid copper cylinder with radius 11.28 cm, and a 1 mm Inconel 718 sheath. This may be viewed as a 'single turn' coil though we acknowledge that in reality a 'multi turn' coil will be needed. Our primary focus here is on assessing the feasibility of copper trim coils to stabilize the plasma as well as survive the high temperature environment.

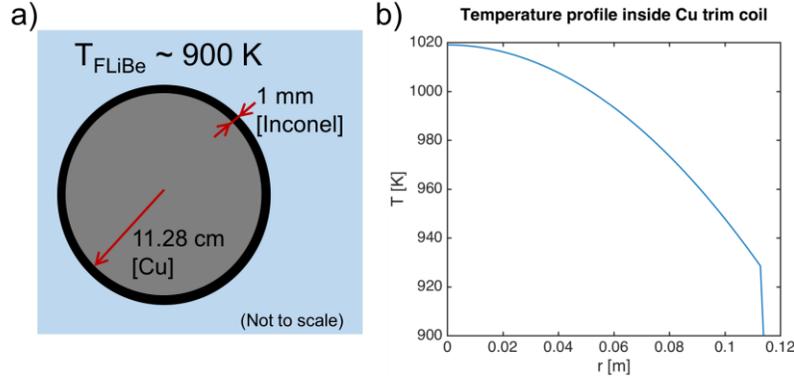

*Figure 3: (a) Cross section of a single-turn copper trim coil used in the present study. This design takes advantage of the electrically insulating properties of the FLiBe blanket, requiring no insulating layer between Inconel and FLiBe. (b) Temperature profile inside the trim coil. Note that the sharp rise at the edge of the coil is due to the poor thermal conductivity of Inconel in comparison to copper.*

Magnetic equilibrium calculations carried out with the modeling software ACCOME [32,33] show that a total current of 20 kA in one single internal coil is able to correct for a 2% change in poloidal flux at the plasma boundary. Therefore, this was taken as the level of current needed for rapid correction to unexpected changes in magnetic equilibrium. Due to the proximity to the plasma, magnetic diffusivity timescales for the trim coils to affect the plasma equilibrium are estimated to be 50 ms (Section 3.4).

Analytical analysis was performed to estimate the ohmic heating and peak steady-state temperatures in these coils that results from running them for an extended period at 20 kA. The calculations were performed assuming the coil is immersed in a bath of 900 K FLiBe with material properties shown in Table 2. Table 3 shows the results of these calculations for one of the vertical stability copper coils. The resulting temperature profile is shown in Figure 3. The maximum on-axis temperature of the coil was found to be 1020 K. The total ohmic heat dissipation in each coil is ~20 kW assuming a worst-case scenario of 10 MW of nuclear heating at the location of the vertical stability coils (see Section 4). This low value of ohmic dissipation is negligible in power balance. Volumetric nuclear heating is the primary cause of the temperature rise inside the coils. When ARC is running, these coils would not operate with DC current, but rather with short pulses for plasma shaping control. However, even for DC operation, the peak on-axis coil temperature appears acceptable.

*Table 2: Thermal and electrical properties of copper, Inconel and FLiBe used for DC heating and analysis and as inputs for COMSOL simulations of magnetic sensing time.*

|  | Thermal conductivity (W/m/K) | Electrical resistivity (Ohm-m) | Melting temperature (K) |
|---|---|---|---|
| Copper | 352 [36] | $6.9 \times 10^{-8}$ [36] | 1,356 [36] |
| Inconel | 22.49 [37] | $1.3225 \times 10^{-6}$ [37] | >1,500 [37] |
| FLiBe | 1 [38] | $6.5 \times 10^{-3}$ [39] | 732 [38] |

Radial and vertical forces acting on the trim coils were estimated by calculating the induced Lorentz forces at 20 kA coil currents in the background magnetic field at the coil locations. These were found to be small (Table 3), and the principal tensile stress in each coil is negligible when compared to the yield stress of Inconel 718 at 900 K.

*Table 3: Operational parameters for single-turn copper trim coils.*

| **Steady state parameters of each in-vessel coil** | **Value** |
|---|---|
| Cross section | 0.04 m$^2$ |
| Total current | 20 kA |
| Ohmic heating (Cu part) | 18 kW |
| Ohmic heating (Inconel 718 sheath) | 0.02 kW |
| Nuclear heating | ~10 MW (MCNP calculations) |
| FLiBe temperature (at contact with Inconel sheath) | 900 K |
| Peak temperature on coil | ~1,020 K |
| Principal stress $\sigma_\theta$ | -2.9 MPa |
| Yield stress of Inconel 718 @ 900 K | ~ 800 MPa [37] |
| Radial force (Lorentz forces at 20 kA coil current) | -0.02 MN/m |
| Vertical force (Lorentz forces at 20 kA coil current) | 0.3 MN/m |

Due to their placement near the vacuum vessel, the vertical stability coils will be subjected to large neutron fluence. However, significant thermal and electrical degradation of the primary copper conductor due to radiation damage is not expected because the operational temperature of the coil is $T/T_{melt} \sim 0.75$. At these high temperatures, most radiation-induced defects will anneal quickly, leading to little long-term damage accumulation [40]. Radiation-induced degradation coupled with FLiBe corrosion of the Inconel sheath will likely be a life-limiting factor for these coils, as more damage will likely be retained at the lower homologous temperature for Inconel ($T/T_{melt} \sim 0.62$). A quantitative estimate of this lifetime is difficult to construct since the coupled effects of radiation and salt corrosion in nickel-based super alloys such as Inconel 718 are potentially important, but not well-explored. However, since these coils are integrated into the VV/divertor assembly, they would be replaced with it, and thus have a short lifetime requirement of ~1 year.

Based on these simple thermal and electromechanical considerations, a copper-based coil design located inside the FLiBe tank and outside the VV appears feasible. Peak temperatures can be maintained below melting temperatures, forces on the coils are small, and ohmic power dissipation is minimal. However, further analysis is clearly needed: assessing plasma stability control requirements with a time-dependent tokamak equilibrium simulation code to determine optimum trim coil locations and currents; coil design considerations for AC operation, accounting for skin effects; electrical insulation of the coil; designing suitable structural supports connecting the trim coils to the vacuum vessel; and developing appropriate electrical connections to outside the FLiBe tank.

### 3.4 Magnetic sensors

The primary challenge associated with the magnetic sensors in ARC, as with sensors in any DT fusion reactor, is balancing the requirement to shield the sensors from the high neutron load (which favors placing them far from the plasma) while maintaining sufficiently fast magnetic response time (which favors placing them close to the plasma). This challenge is particularly difficult when the blanket contains electrically conducting materials, which slows the penetration of the electromagnetic fields and subsequently delays detection of changes to plasma condition at sensor locations. Unfortunately, most materials used for neutron shielding are also electrically conducting (e.g. structural steel, lead, lithium).

In ARC, as with other reactor designs, the magnetic sensors must be placed outside of the neutron shield to reduce the radiation and heat flux exposure of the sensors. The magnetic diffusion challenge, however, is significantly mitigated in ARC because FLiBe, as a molten salt, is only slightly conducting, $\sim 10^4$ times more resistive than a liquid or solid metal (Table 2). Magnetic sensors would consist of flux loops, diamagnetic loops, and Rogowski coils, providing essential measurements of poloidal flux, poloidal magnetic field, and plasma current to implement MHD equilibrium magnetic reconstruction. Non-inductive measurements of magnetic field components (e.g. hall probes) are also needed to compensate for drifts in integrating inductive sensors over long time scales.

In order to assess the detector response time for sudden internal plasma changes and to determine feasible locations for magnetic sensors a model of the full ARC PF coil set, plasma current, VV, and FLiBe tank was implemented in the multi-physics modeling software COMSOL [41]. Time dependent calculations of the magnetic field diffusion due to plasma current change, radial displacement, and vertical displacement were performed. The magnetic diffusion time was measured at the sensor locations shown in Figure 4. External to the FLiBe tank, the sensors would be shielded from neutrons and thermally insulated from the FLiBe tank.

The VV and FLiBe tank were modeled as Inconel 718 and FLiBe having the electrical resistivity properties listed in Table 2. These simulations reveal that a 5 cm vertical or radial step displacements in the last closed flux surface and a 2% step change in plasma current (the maximum allowable change before loss of divertor X-points) can all be detected in approximately 50 ms, assuming magnetic sensors that are sensitive to 0.5% changes in the field. This was deemed a reasonable estimate for sensor sensitivity as tests on JET have found radiation resistant Hall sensors to be sensitive to 0.3% changes [42]. In comparison, a neutron blanket with the equivalent electrical resistivity of Inconel 718 would increase this delayed detection time by an order of magnitude to 500 ms (Figure 5). This is a clear advantage for using a FLiBe tank compared to traditional metallic blankets.

Simulations were also performed to determine the time required for fields generated by internal trim copper coils (Section 3.3) placed inside the FLiBe tank and outside the VV to propagate to the plasma. A 20 kA current step was applied to the trim coils and the propagation time was determined as the characteristic e-folding time for changes to the poloidal magnetic field at the divertor X-point. These coils would be actuated to respond to plasma displacements, allowing control of the divertor X-point locations. The simulations reveal a similar response time (< 50 ms) for the inward propagating fields. Eddy currents in the Inconel portion of the VV and tank determine the field penetration time, while the FLiBe acts essentially as a vacuum to magnetic field penetration. Whether this combined detection and response time (~100 ms) is acceptable for feedback control remains to be determined. If need be, passive stabilizer plates could be considered as an option to reduce active feedback requirements, similar to the ARIES-I proposal [43]. These would be placed around the vacuum vessel, most likely on the FLiBe tank side. The impact on neutron transport and tritium breeding would also need to be assessed.

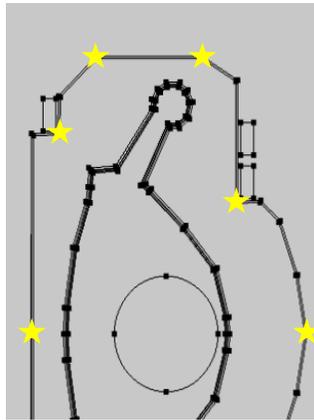

*Figure 4: In order to provide shielding from neutrons and a close to room temperature environment, magnetic sensors are located on the outside of the FLiBe tank. The diffusion times of the poloidal magnetic field perturbations were assessed at the starred locations.*

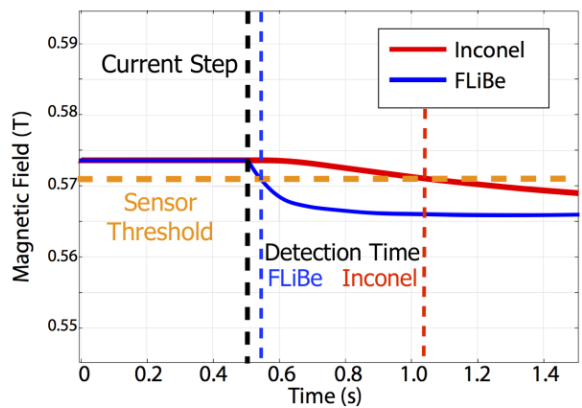

*Figure 5: Poloidal field time history at the outer midplane (a magnetic sensor location) in response to a 2% step change in plasma current. The red trace corresponds to a simulation with the shield tank having an effective resistivity of Inconel 718. The blue trace corresponds to the case of interest – shield tank filled with FLiBe. Assuming a magnetic sensor can detect a 0.5% change in poloidal magnetic field, the low electrical conductivity of the FLiBe blanket allows this change to be detected in under 50 ms.*

## 3.5 Forces on poloidal field coil set

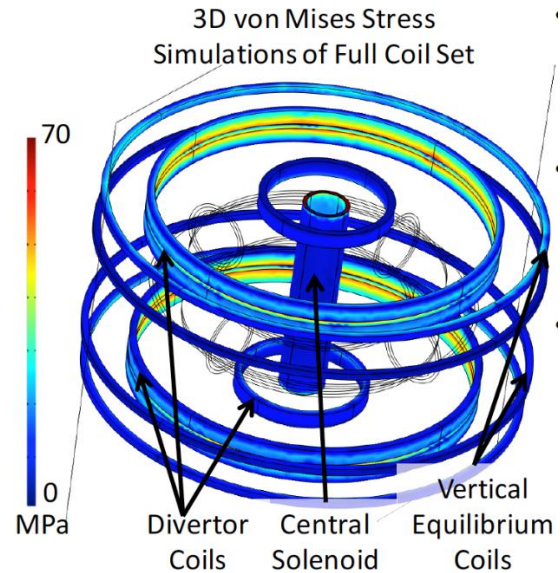

*Figure 6: 3D representation of the poloidal coil set used as the basis of the finite element calculation of Lorentz forces on each coil. Colors reflect the steady state Von Mises stresses calculated as a result of currents applied to all coils. The central solenoid was included in order to assess the effects of the total field on the divertor and vertical coils and was not itself assessed.*

A steady-state analysis of the Lorentz forces on the PF superconducting coils was carried out using the COMSOL multi-physics, finite element modeling software in order to assess requirements for a support structure. Of concern are both the self-induced hoop forces on the coils as well as the vertical forces between stacked sets of coils. Simulations of these forces were carried out using coupled magnetics and structural mechanics packages by prescribing current densities in each of the following coils: the central solenoid, the inner and outer divertor coils, the vertical equilibrium coils, and a circular cross section plasma with a uniform current density. Note that the central solenoid was included primarily to provide the required background magnetic field and its stresses were not evaluated. The specifics of the central solenoid coil were outside the scope of this study but were examined in [1]. The current densities and directions for each of the coils are reported in Table 1 and the imposed plasma current was set to 8.0 MA. Geometric fixed points, necessary for the calculation of static forces, were implemented on the inside bounding surface of each coil to approximate their attachment points to the FLiBe tank. Varying these fixed point locations did not significantly affect the values reported below. The VV geometry was not implemented in these simulations, as it should not affect the steady state coil forces.

Figure 6 shows both the coil set implemented for the Lorentz force calculations as well as the resultant simulated steady state von Mises stress in each of the PF coils. For all coils, the peak stress due to hoop tension was less than 80 MPa, well below the operational yield limit (~800 MPa [37]) of Inconel and of the Hastelloy backbone used in the superconducting tapes. The resulting strain would be ~0.05%, well below the critical axial tensile strain limits for REBCO HTS [31].

The vertical forces on each coil, due to all other coils in the simulation, are shown in Table 4. As the two outer divertor coils (PF2 and PF3) have significant currents in opposite directions, the vertical forces are the greatest on these two coils, pushing them apart vertically. If we assume that these forces are evenly distributed on the poloidal cross sections (6.1 m$^2$) of the coils and the coils are fixed, the PF2 and PF3 coils would produce stresses of 36.6 and 43.0 MPa, respectively, well below the deformation threshold. Therefore, the challenge is simply to immobilize each coil with respect to the FLiBe tank in a reasonable manner. It should be noted that the reduced Lorentz forces is a significant advantage of being able to place the coils inside the TF coils, close to the plasma, as it minimizes the required coil currents.

*Table 4: Vertical Lorentz forces on the poloidal field coil set calculated from COMSOL simulations.*

|  | Divertor coils | | | Vertical field coils | |
| --- | --- | --- | --- | --- | --- |
|  | **PF1** | **PF2** | **PF3** | **PF4** | **PF5** |
| Vertical force (MN) | 20 | -223 | 262 | -76 | 27 |
| Vertical cross section (m$^2$) | 2.3 | 6.1 | 6.1 | 7.7 | 8.3 |

Due to the low vertical Lorentz force and low-level of neutron flux at the coil locations (outside the tank [1]), the vertical equilibrium coils can be supported and attached to the TF coils using standard structural engineering practices. The two outer divertor pull coils could be constructed as a bundle – with a fixed support between and banded together on their exterior. Banding the coils together in this manner allows the bulk of the vertical stress to be compensated in the banding material which would be well within allowable stress limits as mentioned above. While modeling the stresses on these connection configurations is beyond the scope of this work, the manageable Lorentz forces on the poloidal coils suggest that simple engineering solutions should exist for structurally supporting these coils.

## 4. Neutronics

ARC, like all fusion reactor designs, employs a carefully designed blanket that serves the critical functions of neutron shielding, energy capture and tritium breeding. Deuterium-tritium fusion reactions within ARC will produce approximately 2.2×10$^{20}$ 14.1 MeV neutrons per second during full-power operation. These neutrons carry 80% of the energy produced by the DT fusion reaction. Understanding their energy deposition within the VV and coolant is vital to the success of ARC or any deuterium-tritium based fusion power plant. Radiation heating, material damage, and the tritium breeding ratio (TBR) were modeled for ARC using MCNP, a Monte Carlo photon and neutron transport code [44]. To facilitate the development of the MCNP model, coordinates from the simplified VV outline used for the ACCOME magnetic equilibrium scoping study (Figure 1) were taken as the actual first wall shape and converted to a usable MCNP geometry. The VV double wall, beryllium neutron multiplier, tank and neutron shield structures were then added around it. The neutron source location was determined by inner magnetic flux surfaces of the ACCOME magnetic equilibrium. The final design achieves a TBR of 1.08

± 0.004 (Section 4.3), and provides a PF coil lifetime greater than ten full-power years (FPY) (Section 4.2).

## 4.1 Divertor and vacuum vessel design

Figure 7 shows the final geometry as implemented in MCNP. Increasing the length of the divertor leg is found to reduce the high-energy component of the neutron flux in the divertor significantly. This arises from the simple geometric advantage of placing ~1 m of FLiBe along the line-of-sight between the neutron source in the core plasma and the divertor. Figure 8 shows the neutron energy spectrum both at the outer midplane (measured at the plasma facing side of the outer structural Inconel of the VV) and the divertor target volume (measured at the outer surface of the VV at the midplane of the divertor foot feature). The neutron flux in the divertor area is a factor of thirty lower than at the outer midplane. In addition, the neutron spectrum is significantly softened due to the moderation provided by the FLiBe. The reduction of fast neutron flux drastically reduces neutron damage rates, in terms of both displacements per atom (DPA) and helium production from neutron-alpha interactions. Lowering the damage rate in the divertor is highly advantageous for these high heat flux components, as materials subjected to high DPA often show a notable degradation in thermal performance [45]. Materials used in the divertor heat exhaust system must function near their thermal limits (see Section 5). Minimizing this degradation is integral to ARC's long-term operability and, as a pilot plant, demonstrating high availability.

The radial build used in the MCNP model is shown in Figure 9. A non-structural 1 cm thick beryllium layer is placed on the surface of the outer VV wall. This neutron multiplier allows ARC to attain its targeted TBR, as discussed in Section 4.3. For this MCNP assessment, the same radial build shown in Figure 9 was implemented for the entire VV – including the divertor region – even though the divertor target design must be different to handle the higher heat flux (Section 5). Because the divertor is located far from the neutron source, these details do not significantly impact the neutron transport calculations. Total damage values integrated over one FPY for the entire VV for and for just the divertor region are shown in Table 5. The averaged values are high compared to what can be obtained from fast-flux neutron test facilities. However, the integrated damage rates in the divertor region may be obtained in present day fission test facilities. Levels in the range of single DPA per year found in the divertor are on the same scale as commercial fission power systems [46]. This means that materials for such a divertor may be tested under reactor level heat, plasma, and neutron loads without the need for building a high fluence DT neutron source. However, special considerations would still need to be made to account for the higher He/DPA ratio in the ARC divertor.

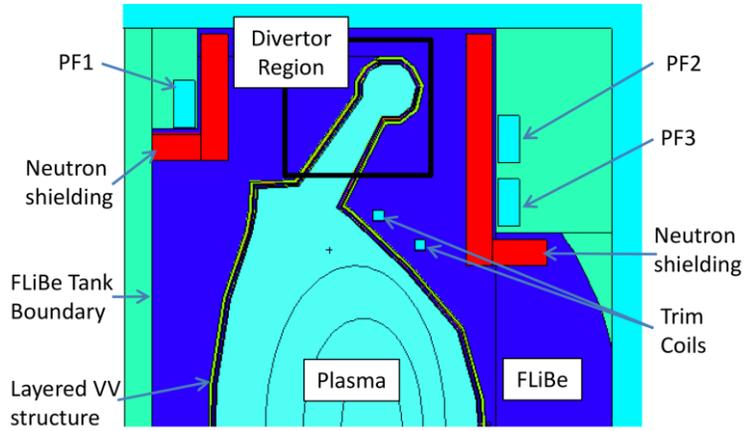

*Figure 7: Cross-section schematic of the ARC design as implemented in the MCNP model. The black box indicates the region used to record divertor neutron damage level tallies listed in Table 5. With 25 cm thick ZrH$_2$ neutron shield plates (red) placed in front of PF1, PF2 and PF3, these coils meet the requirements for full power year lifetime (> 12 FPY, see Table 6).*

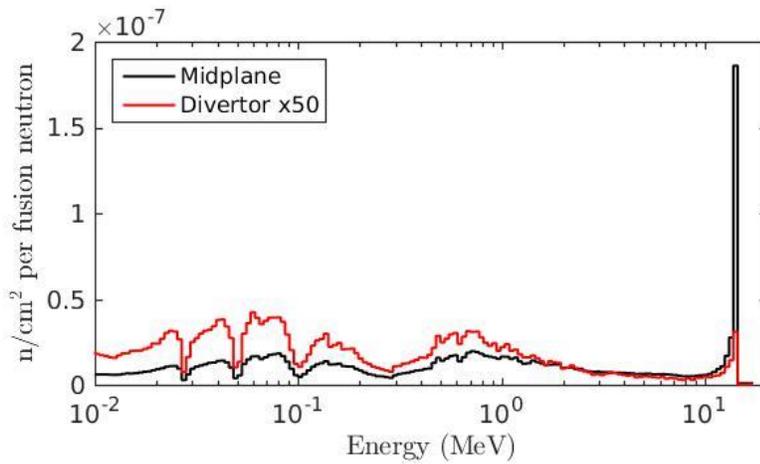

*Figure 8: Shown is a histogram of the neutron energies in the divertor (red) as well as the outer midplane (black). Note that the histogram bins are of variable width, selected to be uniformly spaced on a logarithmic scale. This was chosen to make key features, such as the 14 MeV peak, visible. In this figure, the divertor spectra has been increased by a factor of 50 in order to be visible on the same axis. The neutron energy spectrum at the divertor is significantly softer than at the midplane, limiting the helium buildup through (n,α) processes and lowering the DPA rates on the high heat flux surfaces.*

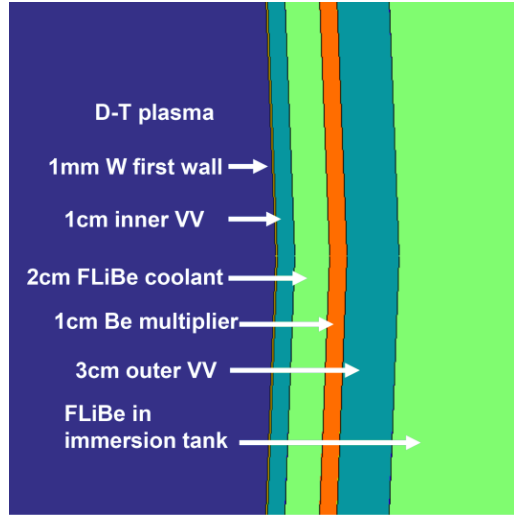

*Figure 9: Radial build of the double wall vacuum vessel implemented in MCNP. A 1cm thick Be layer is placed on the surface of the outer VV layer to function as a neutron multiplier.*

*Table 5: Neutron damage levels for vacuum vessel layers for one full performance year ($P_{fusion}$ = 525 MW) averaged over the entire poloidal cross section and localized to the divertor region (see Fig. 7).*

| Tally location | DPA/yr | He appm | He/DPA |
| --- | --- | --- | --- |
| W inner wall average | 5.4 | 2.4 | 0.45 |
| Inner VV Inconel average | 27.7 | 208.4 | 7.52 |
| Be multiplier average | 9.2 | 2654.6 | 287.7 |
| Outer VV Inconel average | 16.4 | 104.3 | 6.37 |
| Divertor tungsten inner wall | 1.9 | 0.7 | 0.38 |
| Divertor inner VV Inconel | 9.0 | 59.9 | 6.65 |
| Divertor outer VV Inconel | 4.5 | 24.7 | 5.46 |

## 4.2 Neutron shielding

The behavior of the HTS magnets under long-term irradiation, particularly the REBCO tapes used in this design, is not well characterized due to a lack of high-energy neutron exposure data. However, a rough estimate for the lifetime of the HTS superconducting poloidal field coils can be made based on $Nb_3Sn$ data, which is anticipated to have a lower threshold than REBCO [47]. This limit, $3 \times 10^{18}$ n/cm$^2$ (for neutron energies > 0.1 MeV), does not represent the failure point of $Nb_3Sn$, but rather a point at which the critical current begins to degrade. As a result, actual coil lifetimes are anticipated to be longer than the conservative estimates shown (Table 6), in particular because the PF coils are designed to work with a significant margin to their critical current as highlighted in Table 1. The internal trim and vertical stability coils, which are exposed to a higher fluence than the toroidal or poloidal field coils, are copper and therefore do not suffer the same type of degradation. The actual neutron irradiated performance of the HTS used for this design is an open question and should be studied in future work.

The target value for superconducting coil lifetimes in this work is 10 FPYs, consistent with the original ARC design lifetime for the TF coils. It is found that this requirement can be met by adding 25 cm thick shielding plates of zirconium hydride ($ZrH_2$) to the modified VV at PF coil locations. This material reduces both the neutron flux and neutron energy at the coils. The shielding was added on the interior and top/bottom surfaces of the PF coils, as shown in Figure 7. Zirconium hydride was selected because of its high hydrogen density and history of successful use in TRIGA reactors [48]. Other shielding materials may be more desirable based on chemical compatibility and safety. The key takeaway here is that adequately long lifetimes for the PF coils can be obtained by supplementing the liquid FLiBe blanket with moderate amounts of solid target shielding.

*Table 6: Coil lifetimes as calculated based on a $3\times10^{18}$ n/cm$^2$ limit and damage levels from MCNP simulations with and without 25 cm of $ZrH_2$ shielding. Coil lifetimes are presented in full performance years, $P_{fusion}$ = 525 MW.*

| Coil set | Lifetime without shielding (FPY) | Lifetime with shielding (FPY) |
|---|---|---|
| PF1 coil | 2.32 | 12.5 |
| PF2 coil | 6.82 | 76.2 |
| PF3 coil | 1.10 | 11.3 |

## 4.3 Tritium breeding ratio

Tritium is produced within ARC when fast neutrons impact lithium nuclei in the FLiBe blanket. For a viable fusion reactor, more than one tritium atom must be produced per fusion neutron in order to maintain the fuel cycle. With the addition of a continuous 1 cm non-structural beryllium layer as a neutron multiplier (Figure 9), the TBR of the final design is 1.08 (+/- 0.004). This value is slightly lower than the 1.1 TBR reported in the original ARC design [1] and the difference is due to the reduction in FLiBe volume as a result of the addition of the two long leg divertors.

Approximately 26% of the tritium produced within the reactor is generated in the FLiBe cooling channel within the VV wall (Figure 10). In this location, there is a higher flux of fast neutrons relative to outside the VV due to the coolant channel's proximity to the core plasma. Future iterations on the ARC design could potentially increase the TBR by widening this channel.

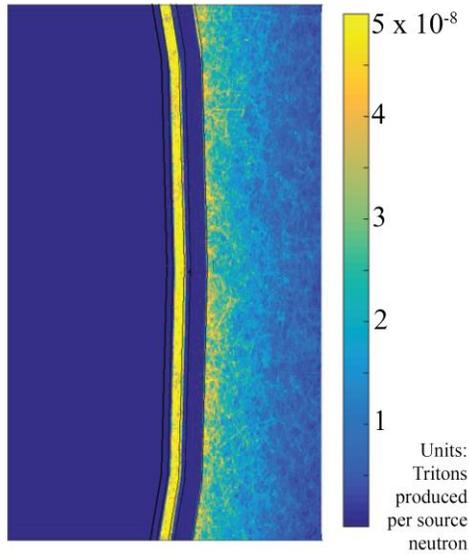

*Figure 10: A large fraction of the tritium produced in ARC is due to production in the coolant channel. Few tritium-producing reactions exist in the structural material of the VV, clearly showing the localization in the coolant channel. The tritium production drops off rapidly with distance from the core plasma as a result of the neutrons being thermalized by the FLiBe blanket.*

### 4.4 Power deposition

Due to the VV's proximity to the core plasma, a nontrivial portion of the neutron power is deposited volumetrically in the first wall, Inner VV Inconel, beryllium multiplier, and VV cooling channels. The distribution of neutron and photon power within the various layers of the reactor vessel are shown in Table 7, based on energy deposition tallies obtained from MCNP. Note that photon power here refers to photons generated from the 14.1 MeV fusion neutrons and not the photons radiated from the plasma. The statistical neutron distribution used to generate these energy deposition estimates is scaled to account for the DT neutron source ($2.2 \times 10^{20}$ s$^{-1}$), which by itself deposits 420 MW fusion neutron power ($P_{fusion}$ = 525 MW) into the blanket materials. The power generated by the blanket exceeds this amount (484 MW), primarily due to exothermic neutron on lithium-6 reactions. (A relatively small but non-negligible number of endothermic neutron capture reactions also occur at high neutron energies, as discussed below.) A primary challenge for the thermal design is that approximately ~30% of this neutron power is deposited into layers of the VV, which must be continuously exhausted into the bulk FLiBe coolant.

It is important to note that the geometry employed by the neutronics model is an approximation to the actual geometry envisioned, particularly in the divertor region. For example, the Be layer in the VV radial build (Figure 9) is taken to be present in the divertor, while in the real implementation of the ARC design, this layer would only be present in the main chamber. Due the softer neutron spectrum in the divertor (see Figure 8), the loss of the Be in the divertor would have minimal impact on TBR or DPA tallies, but energy deposition tallies may be significantly altered due to the moderating properties of Be. The first ARC design study [1] showed that the Be can be replaced by tungsten to act as an effective neutron multiplier – and tungsten is a likely design choice for ARC's divertor targets (Section 5.2). A follow-on MCNP study is needed to do assess the overall impact.

The MCNP simulations and power deposition estimates presented here represent a significant refinement compared to the original study [1] in which the total thermal power generated in ARC was estimated to be 708 MW. The refined calculations project to 630 MW thermal power. This difference is traced to the inclusion of endothermic neutron interactions with elements such as fluorine, which have relatively high threshold energies. These reactions will ultimately produce a mix of isotopes in the blanket and transmutate original atoms to other elements (in the case of fluorine, to oxygen, carbon and nitrogen). The cost for driving this process is a net loss of approximately ~1 MeV per primary 14.1 MeV neutron. This gradual shift over time of the isotopic make-up of FLiBe compared to startup is something that has not yet been analyzed; it would likely impact molten salt chemistry and drive the requirements for continuous monitoring and replacement of FLiBe. TBR could be impacted as well, affecting blanket material choices and design.

*Table 7: Power deposition in each of the material layers modeled in MCNP for 420 MW fusion neutron power. Note that parameters at top are for the entire VV (main chamber and divertors); parameters at bottom are for one divertor only.*

| Layer | Power (MW) | Volume (m$^3$) | Average volumetric heating (MW/m$^3$) |
|---|---|---|---|
| Tungsten inner wall | 8.4 | 0.35 | 24.1 |
| Inner VV Inconel | 39.6 | 3.50 | 11.3 |
| Cooling channel FLiBe | 77.7 | 7.04 | 11.0 |
| Be multiplier | 22.4 | 3.55 | 6.3 |
| Outer VV Inconel | 78.9 | 10.7 | 7.4 |
| Bulk FLiBe | 255 | 241 | 1.1 |
| PF Coil Shielding (see Fig 7) | 1.9 | 49.2 | 0.04 |
| **Divertor Region** | | | |
| Divertor tungsten inner wall | 0.77 | 0.08 | 9.6 |
| Divertor inner VV Inconel | 3.10 | 0.82 | 3.8 |
| Divertor cooling channel | 7.12 | 1.66 | 4.3 |
| Divertor Be multiplier | 1.53 | 0.84 | 1.8 |
| Divertor outer VV Inconel | 5.50 | 2.57 | 2.1 |

## 5. Heat exhaust management system

For the ARC heat exhaust system to be viable, it must ultimately accommodate a wide variety of performance requirements and constraints on materials selection, thermal characteristics, and the pumping power needed for active cooling. Particular attention must be paid to the thermal management of the divertor due to the high heat fluxes that impinge upon those surfaces. As a starting point, we assume that the projected performance of a X-point target, long-leg divertor [21] implemented in the modified VV design (Figure 1) [22], is fully realized in the ARC design. The key attributes are: (1) ability to attain fully detached divertor conditions at ARC's power densities, thereby spreading plasma exhaust heat more or less uniformly over the divertor leg surface area, primarily by radiation, and (2) ability to maintain this passively-stable, fully-detached plasma state over the power exhaust range anticipated for ARC. A high level of plasma control is required as even a brief reattachment of the plasma could potentially devastate plasma facing components. The advantage of the long leg divertor in this regard is discussed in Section 6. Even if the advanced divertor succeeds, one must develop an integrated divertor/VV thermal management scheme. We focus here on one that can take advantage of ARC's unique architecture (e.g. immersion blanket). The goal is to develop and assess a single-point design for the thermal management system and determine its viability based on existing heat transfer technologies and pumping power requirements.

A system-level overview is provided in Section 5.1. We describe the design requirements and constraints, their influence on the resulting system structure, and the overall architecture of the thermal management system for the VV immersed in a bath of molten salt coolant. Sections 5.2 - 5.4 focus on the design and analysis of the divertor cooling scheme, assessing its viability based on existing materials. An assessment of the integrated divertor/VV thermal management system is presented in Sector 5.5. The schemes presented here assume that advanced manufacturing techniques, such as additive manufacturing, can ultimately meet the challenge of producing vacuum tight structures with embedded coolant channels coupled to dissimilar materials. This is obviously a high priority future research need, as discussed further in Section 7.

## 5.1 System level design

In this study of ARC, a flux $2.2 \times 10^{20}$ 14.1 MeV neutrons per second and their interactions in the surrounding structure produce a total of 484 MW of volumetric heating (Table 8). Approximately 227 MW of neutron heating is deposited in the VV (structural material + FLiBe cooling), with the rest deposited in the bulk blanket FLiBe. We assume that the core plasma will radiate ~ 50 MW or 35% of the plasma heating power (alpha power, 105 MW + external heating power, 39 MW) uniformly to the walls of the VV main chamber (area ~ 200 m$^2$). This level of core radiation is consistent with obtaining high performance core plasmas in present experiments. The remainder of the power, 65%, conducts through the SOL and is shared evenly between the two outer divertor legs where it is dissipated by detachment fronts and radiated uniformly over the two divertor 'feet' (see Figure 11). Each divertor chamber must therefore accommodate 47 MW of heat exhaust, appearing as surface heat fluxes (plasma, neutrals, photons) impinging on the walls.

The goal is to identify a design for a heat exhaust management system capable of exhausting this heat while having acceptable pumping power. In addition, material temperatures must remain within standard operational limits to avoid melting and have acceptable structural strain. The design must survive 1 FPY of operation to be consistent with the ARC maintenance schedule [1]. Finally, all electromechanical components (e.g. pumps) must be located outside of the hot FLiBe tank and regions of high magnetic field, and all pipes must enter and exit through the top of the FLiBe tank to allow vertical installation and removal of the VV/divertor/trim-coil assembly.

*Table 8: Summary of ARC power exhaust. Divertor parameters are for both divertors together.*

| Design Parameter | Symbol | Value |
|---|---|---|
| Fusion power (MW) | $P_f$ | 525 |
| Gain | $Q$ | 13.6 |
| **Power Balance** | | |
| Total nuclear heating of VV (MW) | $P_{nuclear}$ | 227 |
| Total nuclear heating of bulk FLiBe (MW) | $P_{FLiBe}$ | 257 |
| Alpha heating power (MW) | $P_\alpha = P_f / 5$ | 105 |
| Auxiliary power (MW) | $P_{Aux} = P_f / Q$ | 38.6 |
| Total thermal power (MW) | $P_{th} = P_{nuclear} + P_{FLiBe} + P_{Aux} + P_\alpha$ | 630 |
| Core radiation fraction | $f_{rad}$ | 0.35 |
| **Divertor Parameter (both divertors included)** | | |
| surface area (m$^2$) | $S$ | 66 |
| power to divertor surfaces (MW) | $P_{div} = (1-f_{rad})(P_{Aux}+P_\alpha)$ | 93 |
| nuclear heating of divertors (MW) | $P_{vol}$ (Table 7: Divertor only) | 36 |
| Average radiative surface heat flux (MW/m$^2$) | $q = P_{div} / S$ | 1.4 |
| Average nuclear volumetric heating (MW/m$^3$) | $\dot{Q}_{vol}$ | 3.0 |

To meet these requirements, a single point design was developed, as illustrated in Figure 11. Pumps are located outside the TF and radiation zone for maintenance. Cool FLiBe at 800 K – providing a margin above freezing (see Table 9) – is pumped into the cooling channels of the VV in four sections: the upper and lower outer divertors, and the inboard and outboard sides of the VV main chamber. Although the four different sections appear to add complexity as compared to a single continuous cooling loop around the VV, this approach is preferred for two reasons: (1) to minimize the total temperature increase across each circuit, and (2) to provide the flexibility to exhaust any increased heat flux due to a slow transient event in specific regions of the VV. The FLiBe simply exits the cooling channel into the surrounding FLiBe tank. FLiBe is then extracted and sent to a heat exchanger to generate electricity. Recirculating FLiBe that is not pumped into the cooling channels is pumped directly into the FLiBe tank and can be diverted to the cooling channels as needed for increased cooling. This allow for the additional capability of variable cooling in the divertor legs without changing the total FLiBe pump rate. The simplicity and robustness of this global heat removal system are evident: due to FLiBe's extremely high boiling point, a single-phase fluid is used without the need for additional pressurization; in addition, all thermal power, including alpha heating, is removed through a single coolant loop.

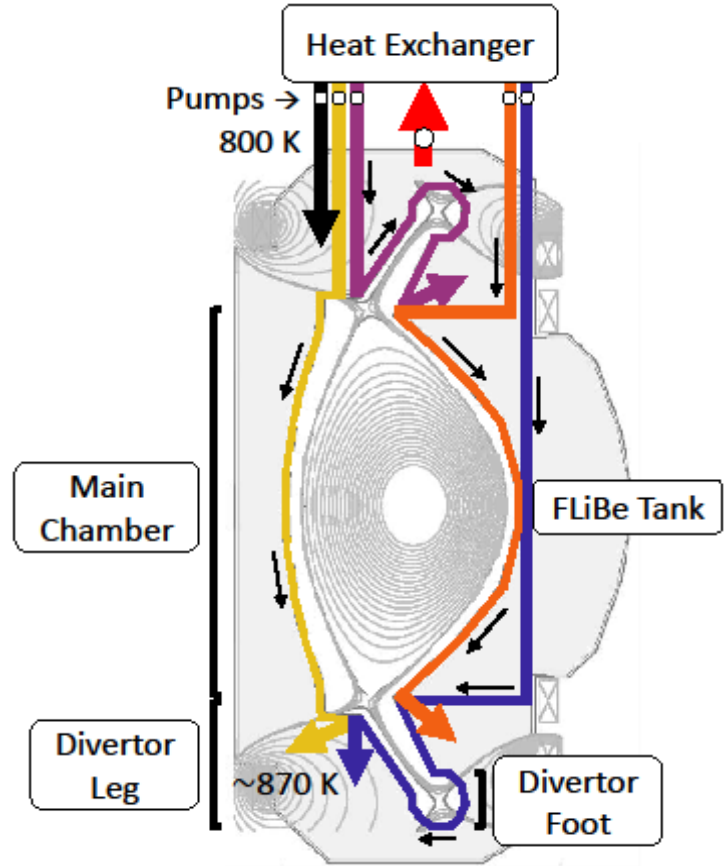

Figure 11: *Schematic of the VV geometry and pumping system. External pumps are represented by white circles. Six representative FLiBe flows are shown with arrows. Note that the pipes and pumps are not to scale, but are meant to convey the FLiBe flow pathways. From left to right, 'cool' 800 K FLiBe is pumped directly into the bulk tank for overflow/reserve (black); around the inboard VV main chamber (yellow); through the upper divertor leg and foot (purple); around the outboard VV main chamber (orange); and through the lower divertor leg/foot (blue). The red upward arrow represents 'hot' FLiBe exhausted to the heat exchanger. Note that all pipes exit through the top of the FLiBe tank to ensure vertical installation and removal of the VV.*

Table 9: *Properties of liquid FLiBe and water at 950 K and 293 K, respectively [1,38].*

| Properties at 1 atm | Symbol | FLiBe | Water |
|---|---|---|---|
| Freezing point (K) | $T_{freeze}$ | 732 | 273 |
| Boiling point (K) | $T_{boil}$ | 1700 | 373 |
| Density (kg/m$^3$) | $\rho$ | 1940 | 1000 |
| Specific heat (kJ/kg/K) | $c_p$ | 2.4 | 4.2 |
| Thermal conductivity (W/m/K) | $k$ | 1 | 0.58 |
| Viscosity (mPa-s) | $\mu$ | 6 | 1 |
| Electrical conductivity ($\Omega$m)$^{-1}$ | $\sigma$ | 251 | 0.005 |

## 5.2 Divertor target design

In some ways, the divertor target plate heat removal challenge for ARC is similar to that facing ITER: the plasma-facing material is tungsten with embedded cooling channels, attached to a structural backbone made of a different material. It is therefore instructive to review the ITER divertor target plate design. The ITER design consists of vertical inner and outer target plates and a central dome to accommodate a single-null magnetic geometry (see Figure 12). The plasma-facing components are tungsten 'monoblocks' with armor thicknesses of 5-8 mm linked by a swirl tube cooling channel [27]. Tiles with 6 mm of tungsten between the plasma and the cooling channels were shown [27,49] to withstand up to 15 MW/m$^2$ steady-state heat flux and higher values transiently. Thus, the embedded swirl tube cooling channel design is highly effective for maximizing heat transfer between monoblock and coolant – a feature that can be employed in ARC. On the other hand, ARC's innovative design results in certain key differences.

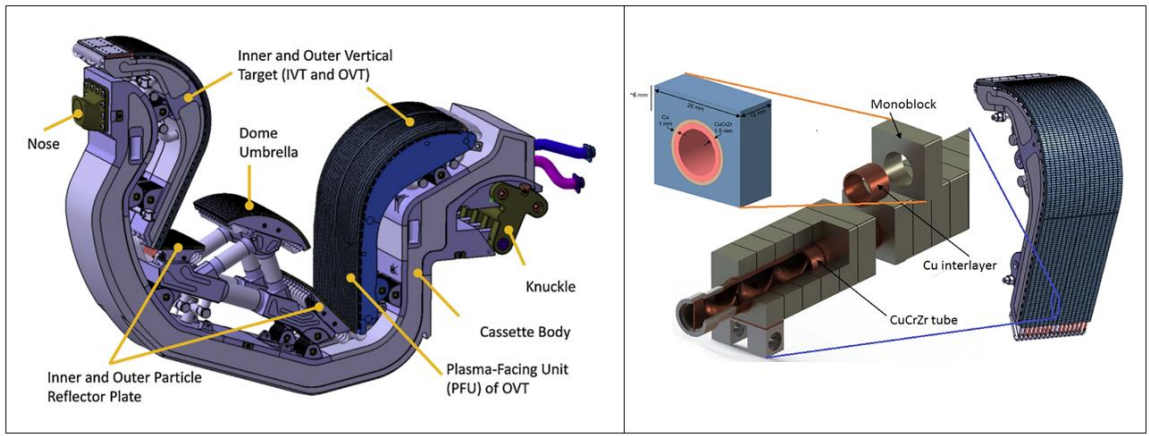

*Figure 12: (Left) ITER Divertor design (figure reproduced from [27], with permission from Elsevier). (Right) ITER tungsten monoblock with swirl tube cooling channel design (figure reproduced from [50], CC-BY-NC-ND 4.0).*

ARC uses a low pressure, high-temperature molten salt coolant instead of high-pressure water. ARC's vertical maintenance scheme, which allows the VV/divertor assembly to be removed as a single unit, is potentially game changing. The need for remote maintenance in the tight radioactive confines of the vacuum vessel in a fusion reactor has led to the development of large robotic arms that are slow and unwieldy. Having to stop operations for lengthy periods of time just to perform minor repairs could be costly. For example the need to cut and re-weld coolant channels, and certify them to nuclear standards, could be prohibitive. In comparison, being able to remove the entire VV/divertor assembly to perform repairs and assessments ex situ provides the opportunity to replace the entire assembly if needed, thus potentially reducing overall downtime and also reducing component lifetime requirements. The expected cost and lifetime of the VV is still under investigation as well as the estimated time required to replace the VV. The exact level of improvement that the vertical maintenance scheme provides compared to conventional approaches is yet to be determined. Finally, the long-leg divertor significantly reduces the divertor target plate heat flux density in ARC compared to standard vertical target divertors whilst a stably detached divertor plasma virtually eliminates plasma-induced divertor target plate material erosion since the incident ion energy falls below the sputtering threshold.

ARC is designed to operate in a double-null divertor configuration and employ a tightly-baffled, long-leg outer divertor geometry (Figure 1) with an embedded divertor X-point. The leg is approximately 1.5 m long with the secondary divertor null centered in a 'divertor foot' (the circular section at the end of the long divertor channels Figure 11). The foot region shape is a partially closed (in the poloidal plane) torus with minor radius ~0.5 m and major radius of 3.7 m. The radially extended leg: (1) expands total magnetic flux by a factor of 1.3 (due to larger R than primary X-point) in order to promote stable divertor detachment, (2) provides large surface area for heat deposition (assumed to be primarily via photon, neutrals, and cold plasma recycling on the side walls from a detached plasma state), and (3) significantly shields divertor surfaces from core plasma neutrons. These features seek to separate the plasma-facing component design challenges of neutron heating and damage, plasma erosion, and high heat flux management. Further optimization of divertor chamber surfaces could be performed. For example, in the divertor foot, tilting target plates at the three strike point locations would allow for better handling of power loads during an event of plasma reattachment, but the simplified geometry shown here is adequate for this scoping study. In addition, there are significant advantages for stability and control to operating with the detachment front located along the leg as will be discussed in Section 6, the leg shape would also then be optimized to minimize surface heat loads accordingly. For simplicity, it is assumed that the plasma detachment front is highly localized at the secondary divertor X-point, centered in the divertor foot. Radiation heat flux is assumed uniform over the torus-shaped divertor foot. In this case, the resulting divertor wall heat flux density is 1.4 MW/m$^2$. Despite this low value, the design specification was set to 12 MW/m$^2$ for the plasma facing components in the divertor leg and foot due to uncertainty in the distribution of the heat flux and to provide a large margin of safety.

ARC's design also introduces new challenges – corrosion of coolant channels by FLiBe and very high coolant temperatures, 800 K FLiBe compared to ~390 K water. Both of these impact the allowed thickness of the first wall, which has plasma on one side and coolant on the other.

While thermal requirements set a maximum wall thickness, plasma-induced erosion and chemical corrosion are important in determining the minimum first wall thickness. Plasma-induced erosion of divertor target materials is dominated by physical/chemical sputtering and possible melting from transient events like ELMs [51-53]. By operating in a stable fully-detached divertor regime, erosion rates can be reduced to negligible levels especially for high-Z materials [54-56]. The design therefore assumes that chemically compatible medium-to-high-Z materials, operating far from their melting points, will experience small levels of net erosion (< 1 mm/year) on the plasma-facing side. On the coolant side, flowing molten salt leads to material corrosion, which not only wears away material surfaces but also causes impurity uptake in the FLiBe. Currently, little data exist on flowing FLiBe corrosion rates. However, studies have shown that for static FLiBe at 873 K, both Inconel 600 and 625 erode at rates of 2.8 µm/year and 1.1µm/year, respectively [57]. The materials in ARC will have thicknesses on the order of millimeters to centimeters and operate at similar temperatures, but flow velocities of several meters per second will likely enhance corrosion. Progress is underway to measure these effects [58]. For the present analysis, a minimum first wall thickness of 3 mm was chosen to accommodate ~1 FPY of operation.

With these considerations in mind, the following divertor target design is proposed: The plasma-facing first wall is 3 mm solid tungsten, actively cooled by FLiBe flowing at 2 m/s through 12 mm diameter swirl tube channels. An Inconel-718 backbone (4 cm thick) provides structural support. This thickness was chosen to be consistent with disruption mechanical stress analysis in the original ARC paper [1] which was not performed in this study. Note that the original ARC study found that the safe dissipation of thermal energy in the event of a disruption remains a challenge. An idealized cross-section of the divertor walls is shown in Figure 13.

As discussed in Section 5.3-5.5, this arrangement is found capable of exhausting an incident surface heat flux of 12 MW/m$^2$ and volumetric neutron heating consistent with Table 7. The difference in temperature at the inlet and outlet of each cooling channel is limited to $\Delta T \leq 75$ K, chosen to keep material temperatures and stresses below their operational limits. After passing through the divertor leg section of the VV, the FLiBe exits into the bulk FLiBe tank where it is then exhausted to a heat exchanger at a final outlet temperature of ~910 K.

This single-point design analysis considers only two divertor materials: tungsten and Inconel-718. Tungsten, which has been used in several divertor designs, was chosen due to its high thermal conductivity, high operating temperature window (up to ~1500 K, its recrystallization temperature), and resistance to plasma erosion. While tungsten at room temperature is brittle, the molten FLiBe cooling system (~800-900 K) will keep tungsten temperatures above the ductile-brittle transition temperature throughout the entire divertor. This is expected to ease mechanical concerns about a pure tungsten divertor.

Inconel was chosen as a representative structural material due to its strength at high temperatures, relatively low electrical conductivity, and familiarity to the fusion community. However, it is recognized that use of a high nickel content material would result in significant radioactivation over the lifespan of the vessel, the potential impact of which has just begun to be assessed. An assessment of material interface issues (thermal conductivity, attachment, differential thermal expansion) and the identification of methods to manufacture the divertor geometry (a highly-simplified version is shown in Figure 13) are also beyond the scope of this study. These are areas in which the development of new materials and manufacturing techniques, such as additive manufacturing, will ultimately determine if a viable approach exits. Due to inconsistent data on high-temperature irradiation of these materials, un-irradiated material properties were used throughout this analysis.

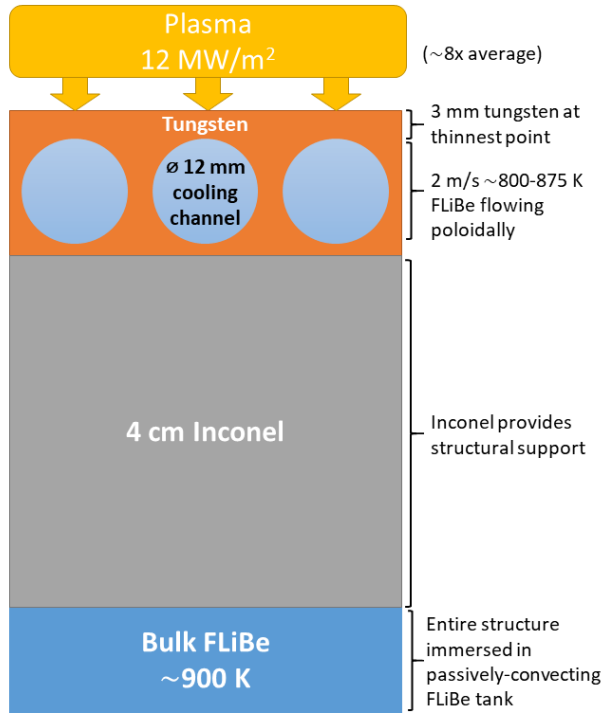

*Figure 13: Cross-section of divertor material layering. The first wall is modeled as being continuous. Swirl tapes are used in the coolant channels (not shown).*

## 5.3 Divertor thermal analysis

To characterize the active cooling in the divertor, a swirl tube design similar to the ITER divertor swirl tube [59] was studied, with an internal diameter of 12 mm and a 0.4 mm thick, 20 mm pitch internal swirl tape. This geometry was modeled with COMSOL using a coupled turbulent k-ε fluid model and temperature-dependent FLiBe properties from [60]. Simulations were repeated, varying the uniform normal inlet velocity from 1-6 m/s to optimize high turbulent heat transfer and minimize total pressure drop which affects the pumping power required.

An effective convective heat transfer coefficient $h_{eff} = q/\Delta T$ was calculated from the applied heat flux and the difference between two temperatures: (1) the average channel wall temperature (nearest the first wall) and (2) the free-stream flow temperature (farthest from the first wall). This value was then used in further analysis to calculate the pressure gradient (Table 10). Temperature and pressure profiles at 2 m/s inlet flow are shown in Figure 14.

*Table 10: Results for swirl tube (divertor; inlet area 113 $mm^2$) at 12 $MW/m^2$ surface heat flux and 2 m/s FLiBe coolant flow.*

| Geometry | $T_{channel\ wall}$ (K) | $h_{eff}$ (kW/$m^2$/K) | dP/dx (kPa/m) |
|---|---|---|---|
| Swirl Tube | 856 | 233 | 350 |

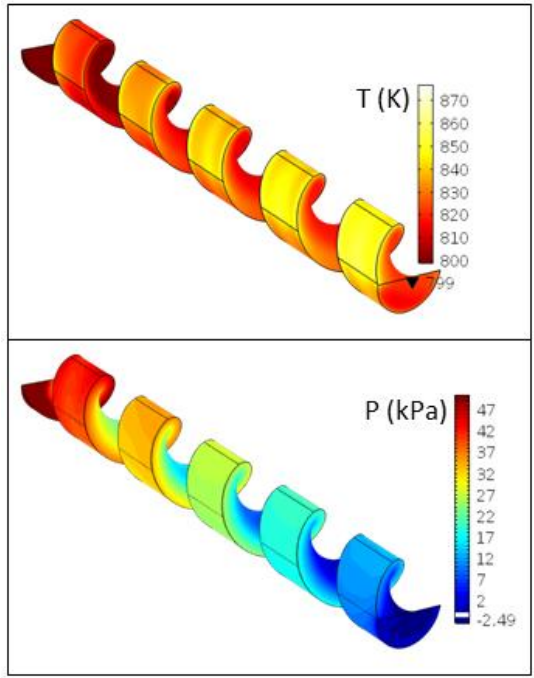

*Figure 14: COMSOL simulation results for temperature [K] (top) and pressure [kPa] (bottom) for the swirl tube. FLiBe coolant flow speed is 2 m/s.*

From [38], it is expected that heat transfer in FLiBe will be affected by MHD effects, but pressure drop will not. Nakaharai et al. [61] report that MHD effects can reduce the Nusselt number and therefore convective heat transfer coefficient by as much as 24% for a FLiBe simulant. For the FLiBe parameters considered in this study, our analysis conservatively reduces the convective heat transfer coefficient by 30%, which further increases the plasma facing wall temperatures.

### 5.4 Divertor thermomechanical analysis

To assess the viability of the design with regard to steady state, thermal induced stresses, the results from Section 5.3 were used as inputs to a thermal stress model in COMSOL. In order to save computational time a single trapezoidal prism cooling channel section was simulated. The non-parallel faces were tilted to match the curvature of the divertor walls. Symmetry conditions were applied on these tiled faces, which are internal to the divertor, by allowing no heat transfer or expansion out of them. The top and bottom faces, which correspond to the inner plasma-facing side and outer bulk tank facing side respectively, were both free to expand in their respective normal directions. Temperature-dependent material properties from [62] were used.

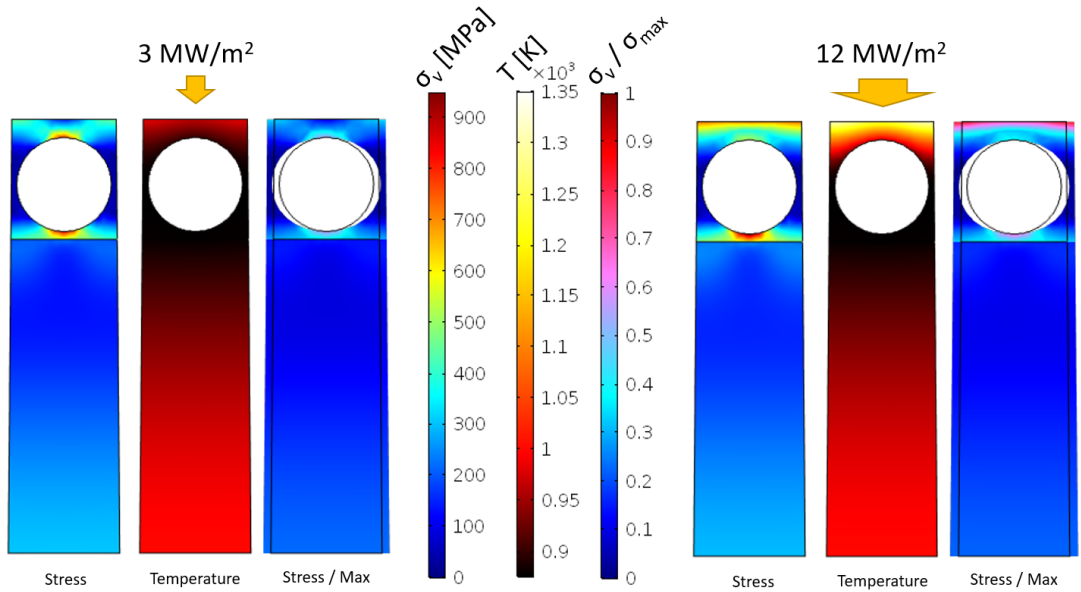

*Figure 15: COMSOL thermomechanical simulation results at two different heat fluxes, (left) q= 3 MW/m² and (right) q= 12 MW/m². The plots under each condition, from left to right, are the resultant von Mises stress [MPa], temperature [K], and the ratio of von Mises stress to maximum allowable stress. The peak temperatures are 1000 K and 1320 K, respectively, and peak stresses are 866 and 934 MPa, respectively, for the different heat flux cases.*

Heat transfer through the channel walls was modeled using a FLiBe temperature of 875 K in the cooling channels and convective heat transfer coefficient of $h_{eff}$ = 163 kW/m²/K. A pressure of 1.4 MPa was applied normal to the walls of the coolant channel to simulate the fluid pressure applied by the FLiBe in the channel on the side walls. This pressure was calculated for a cooling channel located at the bottom of the vacuum vessel at the outlet of the cooling channel loop where the combination of pumping pressure and pressure due to gravity was expected to be the highest. Two separate target heat flux scenarios of 12 MW/m² and 3 MW/m² were applied normal to the plasma facing surface. Average volumetric heating in each layer from the neutron fluxes were included (see Table 7). The outer Inconel face was held fixed at 1000 K consistent with the original ARC simulations [1].

Based on Figure 15, the tungsten layer has peak temperatures of around ~1320 K for the 12 MW/m² case, which is below the recrystallization temperature of ~1500 K. The resulting von Mises stress distribution is also shown, with a peak stress of 934 MPa. This stress was found to be primarily due to secondary stress from thermal loading and geometric discontinuities. Primary membrane stress, introduced around the edge of the FLiBe channel by the pressure of the fluid, was found to be negligible in the point design. The FLiBe pressure in our design contributed only about 30 MPa of stress near the channel walls, roughly 3% of the peak stress there.

Most of the stress resulted from either induced internal thermal gradients from the high heat flux in the 12 MW/m² case, or the geometric concentration of stress induced by differential thermal expansion of tungsten and Inconel. The contribution of thermal gradient stress and differential expansion can be seen by comparing the 3 MW/m² and 12 MW/m² cases, and observing that the areas of peak stress at the plasma-facing wall and at the material interface are greatly reduced under reduced heat fluxes. The high stress region above and below the cooling channel are due to the deformation of the channel geometry from the toroidal expansion of the entire vacuum vessel with the increase in temperature. All these stresses classify as secondary stresses, as yielding of the material will reduce the induced stresses.

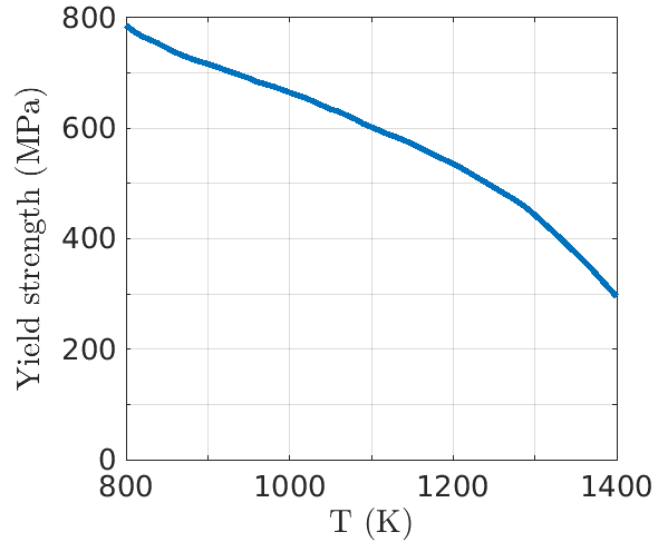

*Figure 16: The yield strength of tungsten varies significantly over the temperature range expected in the divertor, data from* [63].

Using the criteria set by the American Society of Mechanical Engineers (ASME) [64], the maximum secondary stress safely allowed in a material is $\sigma_{max} = 2\sigma_{yield}$, where $\sigma_{yield}$ is the minimum specified yield strength. Note that the temperature-dependent yield strength of tungsten, as seen in Figure 16 [63], was taken into account. Shown in Figure 15 is the ratio of von Mises Stress to $\sigma_{max}$, showing that the stress throughout the model is lower than our maximum allowable stress criterion. Several design options exist to alleviate this stress even further, such as modifying the channel geometry to allow for thermal expansion and to reduce stress concentrations. In addition, an appropriate buffer material or a graded material boundary could be used at the tungsten-Inconel interface. Tungsten/copper grades have been successfully manufactured for decades with the intention of reducing thermal stresses [65]. Further research could be pursued to explore and improve the performance of other tungsten graded materials. Note that the stress limits identified here are a necessary but not sufficient requirement as cyclic fatigue could be a problem. However, as this relates to the expected lifetime of the VV/divertor assembly components, it is outside the scope of this study and will be a topic of future research.

Based on the current design, the peak allowable temperature rise of the FLiBe in the cooling channels was limited to 75 K to ensure that all stresses and temperatures were below allowable limits. Optimizing the design to reduce stress concentrations could raise this cap but such an optimization was not included in this study.

## 5.5 System level analysis

In this section, we investigate the fluid temperatures, flow rates, and pumping power required to exhaust the total heat of ARC. The development and analysis of a cooling channel design for the VV main chamber is outside the scope of this paper. Therefore, for the purpose of a system analysis, we assume that it can attain a similar level of performance in terms of heat removal and pressure drop as the swirl-tube design adopted for the divertor. In addition, for simplicity the allowed increase in temperature through the VV main chamber cooling channels is also set to 75 K.

Given the total heating power to be exhausted and the allowable coolant temperature increases, the required volumetric flow rate of coolant is easily calculated (Table 11). The cool 800 K FLiBe will be pumped externally through Inconel 718 pipes (walls 10 mm thick) that enter and exit the blanket tank vertically in between the 18 toroidal field coils. Flow velocity in these pipes is assumed to be ~2 m/s, and the total area required by the pipes is ~5% of the total useable tank area between the TF coils (~1.5 $m^2$ of ~30 $m^2$), allowing significant space for mechanical supports, vacuum pipes, RF waveguides, diagnostic feedthroughs, etc. [1]

*Table 11: Design specifications for the total heating power to be exhausted and the resulting required volumetric flow rates for an increase in cooling channel temperature of ΔT = 75 K.*

| VV Section | Total Exhausted Power (MW) | Total Required $\dot{V}$ ($m^3$/s) |
|---|---|---|
| Divertors | 129 | 0.37 |
| Main Chamber | 241 | 0.69 |

With the goal of allowing for variable cooling in the divertor legs without changing total FLiBe volumetric flow rates through the system in normal steady-state operation, an additional 0.19 $m^3$/s of FLiBe is pumped directly to the FLiBe tank. This provides the ability to double the cooling channel flow rate, and thus increase the heat exhausted, in one of the two divertors during increased heat loading; for instance, this could occur if a vertical displacement of the core plasma causes unequal sharing of the heat exhaust between the two divertors. Based on this design, the final exit temperature of the FLiBe exhausted to the heat exchanger is 908 K. This is acceptable relative to Inconel's peak operating temperature (~1000 K), leaving some margin for local hot spots in the bulk tank where the FLiBe may be relatively stagnant.

The pumping power required to exhaust the heat from ARC is equal to the product of the volumetric flow rates and the pressure differentials across the various pipes. When considering the change in pressure throughout the entire system, the flow circuit can be split into three main sections: FLiBe moving through external pipes to the cooling channels, through the cooling channels to the blanket tank, and from the blanket tank out of the top of the machine to the heat exchanger. The pressure drops were calculated by taking into account each segment's length and corresponding pressure gradient. For the pipes, the pressure drop was calculated as $\Delta p = k\rho v^2/2$, where ρ is the mass density of FLiBe, and v is the flow velocity; the factor k incorporates effects of pipe bends, expansion, and friction using the Colebrook-White equation [66] which assumes turbulent flow and an Inconel surface roughness of 10 µm [67]; for all cooling channels, the value from COMSOL (Table 10) was used. The required pumping powers are listed in Table 12. The total pumping power of ~3.1 MW is less than 1% of the total fusion power. Note that efficiencies and pumping power through the heat exchanger and external pipe networks have not been included in this analysis.

In summary, we find that the presented heat exhaust system would remove the total power deposited in the VV and FLiBe tank with reasonable pumping power requirements and temperature margins. The system relies on a robust divertor design that is capable of spreading the plasma exhaust heat over a large surface area. A full plant model of the flow of FLiBe through the VV and FLiBe tank is needed to more accurately assess temperature profiles and required pumping power. However, such models are computationally expensive due to their scale and beyond the scope of this paper.

*Table 12: Required pumping powers for separate components of the overall pumping system.*

| Pumping system section | Pumping power required (MW) |
|---|---|
| Pipes to cooling channels | 0.02 |
| VV main chamber cooling channels | 2.50 |
| Divertor cooling channels | 0.42 |
| Bulk tank to heat exchanger | 0.11 |
| **Total** | **3.05** |

# 6. Divertor detachment control and novel diagnostics enabled by the unique design features of ARC

Diagnostics in a reactor will provide feedback control and monitoring of plasma and divertor conditions, which are more challenging tasks in a reactor than in present-day experiments and ITER. The intense neutron environment severely limits the kinds of diagnostics that can be employed; furthermore, the need to maximize the neutron power captured for energy generation places tight space constraints on all planned diagnostics. The heat exhaust system introduced in this paper will require feedback detachment control, some of which can be provided passively on a fast time scale through the long-legged divertor, and some of which can be provided on a slower time scale through a microwave reflectometry/interferometry diagnostic enabled by the unique geometry of the long leg. This section discusses the need for a passively stable detached divertor in a fusion reactor and the novel diagnostics enabled ARC reactor design that could be implemented in conjunction with more traditional diagnostics. We note that the ARC design will need a separate requirements analysis to address the overall minimal viable diagnostic set.

## 6.1 The need for passively stable detached divertors

Divertor heat flux control for a fusion reactor is a very active and challenging area of research. The primary challenge is to ensure that a detached divertor state is maintained at all times, avoiding the focussing of power onto narrow 'strike-point' regions on divertor wall surfaces. To this end, one may be tempted to 'over-mitigate' the heat flux by, for example, injecting seed impurities to enhance divertor radiation beyond what is needed to accommodate the heat exhaust. However, experiments have shown that 'over-mitigation' of the heat flux can result in a MARFE [30] at the main plasma X-points, which leads to degradation of the core plasma confinement [68,69]. In addition, seeded impurities can reduce core fusion reactivity. Thus the 'detachment power window' is found to be very narrow in present experiments that operate with conventional divertors [70]. In this situation, active control of divertor conditions is required to maintain the detached state, such as via a feedback control of seed impurities. Experiments performed to-date have used a variety of divertor diagnostics to perform this sensing and control function: Langmuir probes [71], surface thermocouples [72], vacuum ultraviolet spectroscopy [73], tile current shunts [74-76]. These experiments have been fairly successful in demonstrating a control system that can accommodate steady-state plasma conditions. However, these systems have two main limitations that prohibit their practical implementation in a reactor.

First, the diagnostics currently used are not able to tolerate the high neutron flux environment of a pilot plant. Refractive components, including vacuum windows, experience radiation-induced absorption, which causes opaqueness, and radiation-induced emission, which leads to anomalous signals [77]. Mirrors are vulnerable to damage from plasma surface interactions (erosion and re-deposition), while electrical components and insulators will be damaged by neutrons, experiencing radiation-induced conductivity, electrical degradation, and electromotive forces [78]. In addition, ARC's VV and surrounding FLiBe will be at temperatures much higher than those at which present diagnostic systems are designed to work and the total areal access to the plasma is small because ARC does not have any horizontal ports between the TF coils.

Second, divertor feedback control systems presently employed are unable to cope with power exhaust transients that can occur on millisecond time scales [79]. For example, ARC can experience prompt changes in core plasma confinement modes (e.g. H-L transition). This could result from RF irregularities, accidental impurity injections, or other unplanned variations in device operation. The sudden release of energy into the scrape-off layer would cause the divertor plasma to reattach in less than 1 ms. These timescales are much faster than system response times of any currently available feedback system actuator [79]. Basic considerations such as the volume of the divertor and the rate in which impurity seed gasses can be delivered to the divertor limit feedback control system response times to ~100 ms or longer [50,80]. The intense surface heat flux resulting from plasma reattachment at the strike point is calculated to be on the order of 170 MW/m$^2$ in ARC, akin to a plasma torch. Clearly, this situation must be avoided.

In light of these considerations, the ability to implement a long-leg, X-point target divertor geometry into the ARC pilot plant design with no impact on core plasma volume (Section 3) and acceptable neutronics (shielding, TBR – Section 4) is transformative. Theoretical studies have shown that long legged divertor geometries provide significant advantages to the stability of a detached divertor to transient disturbances in the heat flux [34,81]. Computational simulations of long legged divertors have also been performed [21,82], indicating large enhancements in performance. Figure 17 shows results from the UEDGE study of ADX [21] where four different divertor geometries were compared under otherwise identical core plasma conditions. A scan of the power flow into the scrape-off layer showed that the X-point target divertor geometry (XPTD) maintains a passively stable, fully detached divertor condition at significantly increased exhaust power – a factor of 10 higher than conventional divertors (e.g. SVPD). As exhaust power is varied, the position of the detachment front in the leg self-adjusts as needed in response to the power exhaust – accommodating a factor of 10 variation.

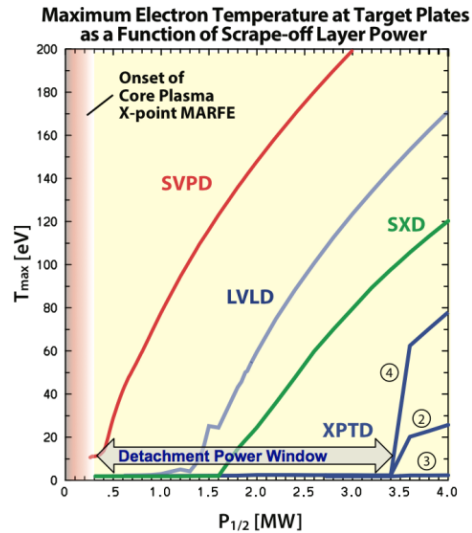

*Figure 17: Maximum temperature on the outer divertor target plotted versus input power into the lower half of a double-null plasma for: standard vertical plate divertor (SVPD); long vertical leg divertor (LVLD); super-X divertor (SXD); XPT divertor (XPTD). Plasma detachment is identified as when the peak target temperature goes below 5 eV. The shaded region at low input power corresponds to the onset of a core plasma x-point MARFE. The power window over which a passively stable detached divertor condition is achieved increases significantly for long-leg divertors, approaching a factor of 10 for the XPTD concept as labelled in the figure. Adapted from [21], with permission of authors and AIP Publishing.*

These results highlight three important advantages for using a long legged divertor in ARC, and in particular an XPTD: as a fusion reactor power exhaust system, this system accommodates the highest power exhaust density; it provides the largest power window for attaining a passively-stable detached divertor; and the location of the detachment front position in the leg can be used as a means for sensing the power load and divertor response (Figure 18a). Since the detachment front is passively stable and can respond immediately to fast changes in power exhaust, there is no need for a fast feedback system to actively attain and control divertor detachment. Instead, focus can be placed on developing neutron-tolerant diagnostics to monitor the detachment front location along the divertor leg and the distribution of heat in the divertor legs at strategic points over long time scales (~ seconds). On these time scales, the location of detachment front in both the upper and lower outer divertor legs could be adjusted such that they are at the nominal operating point – in the center of their respective power windows. Based on modelling results shown in Figure 17, this would allow the ARC divertor to passively and promptly accommodate a ±85% variation in exhaust power. Long time-scale feedback adjustments include: power levels of RF heating systems; main plasma upper/lower X-point flux balance to share power among the divertor legs; changing the impurity seeding levels; changing neutral pressures in the divertors via controlled bypass leaks [83] and/or the rate or location of gas injection for fuelling.

## 6.2 Monitoring divertor detachment location with microwaves

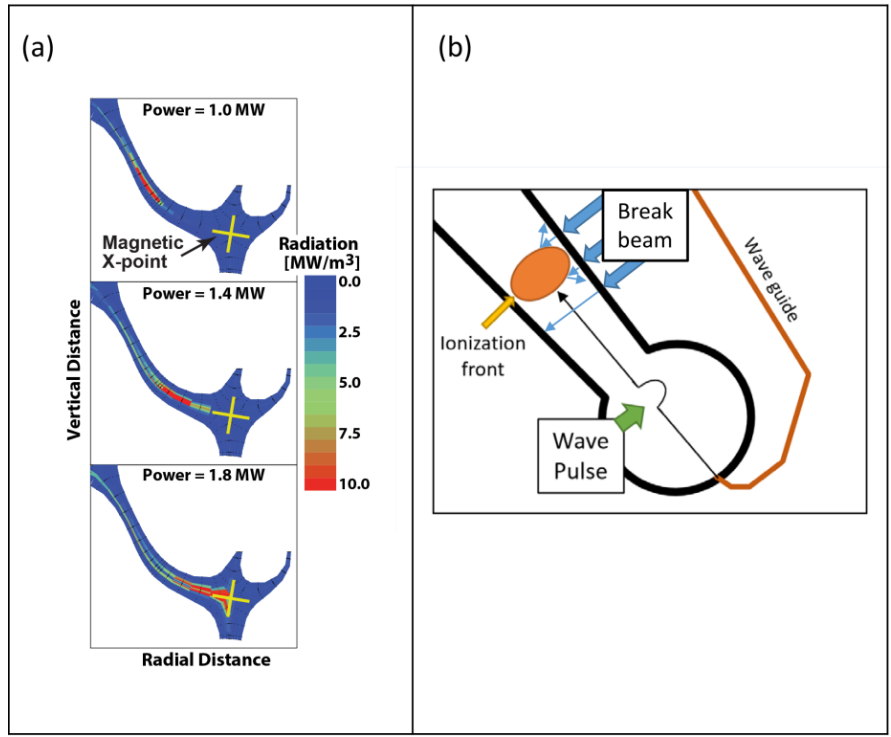

*Figure 18: (a) Simulations show that the detachment front is stable in the long legged divertor and its location along the leg is dependent on the power into the leg [21]. (b) With the use of microwave based diagnostics employing neutron-tolerate metal waveguides, it would then be possible to detect the location of the 'density step' at the detachment front and control the power exhaust to the divertor and/or divertor conditions (via impurity seeding, neutral pressure control, etc.) such that the detachment front is located at its nominal operating point. This would provide the largest margin to handle power exhaust transients while maintaining a detached plasma state.*

A high-density plasma is formed at the detachment front location; further downstream of this location, the plasma density drops precipitously. This makes it ideal for detection using a microwave reflectometry/interferometry system. Metal waveguides are neutron and plasma erosion/redeposition-tolerant and can be used to send microwaves through the FLiBe tank, keeping electronics shielded and protected from neutron damage outside the TF volume. Vacuum windows can also be placed in low neutron fluence zones, ensuring a long lifetime. Similar systems have been explored in DIII-D [84], JET [85], and ITER [86,87]. This precedent gives us confidence that it would be feasible for our applications.

Two possible variants to such a system are: a 'break beam' system and a 'wave pulse' system (Figure 18b). The 'break beam' would rely on measurements of wave transmission perpendicular to the divertor leg at regular intervals along the length of the leg. The reflectometry frequency is chosen to match the cutoff frequency of plasma close to the detachment front (e.g. for plasma densities $\approx 10^{21}$ m$^3$, cutoff frequency $\approx 285$

GHz). The "break beam" technique would provide a binary measurement that can be used to infer the location of the detachment front. The proposed system is intentionally simple since signal degradation in long waveguides may make interpretation difficult.

In comparison, the 'wave pulse' system would have to rely on a finely tuned reflectometry based diagnostic that measures the time-of-flight of a wave pulse. The reflectometry waveguide would enter the divertor at the end target and be aimed up the divertor leg. This would allow for a measurement of the distance of the ionization front from the divertor target. Tests of time of flight measurements, based on a pulse reflected off a plasma, have been made and have been shown to achieve good time (2.5 µs) and spatial resolution (6 mm) [88]. However, such a measurement would be more delicate and care would have to be taken on the calibration and interpretation of the results.

## 6.3 Infrared imaging of divertor 'hot spots' through the FLiBe blanket

One way to monitor the heat flux distribution in the divertor is to measure the surface temperature of the divertor. To obtain this information, ARC could use a set of 'hotspots' that poloidally ring the divertor, as shown in Figure 19.

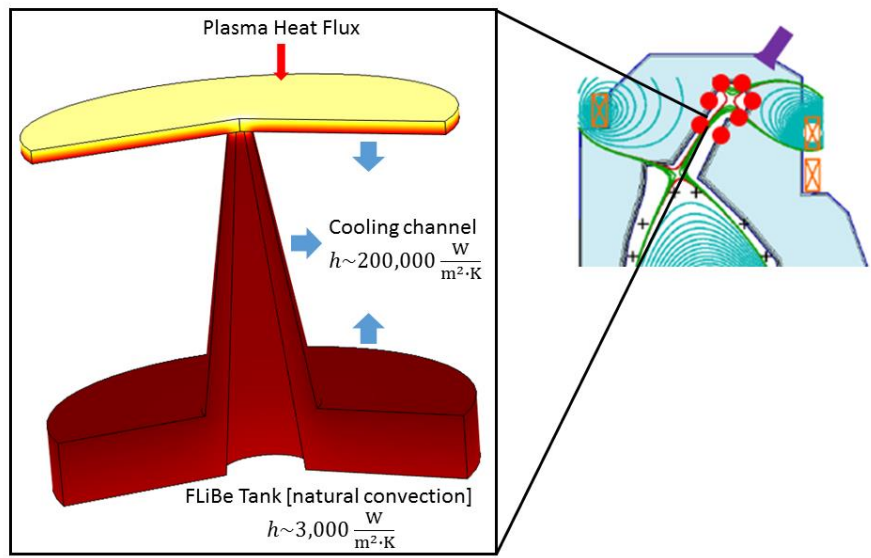

*Figure 19: Hotspot configuration for divertor thermal sensing. They would be located poloidally around the divertor leg to provide a measurement of divertor surface temperatures. The red dots shown indicate these hotspot locations that would be viewed from the neutron shielded optics (purple camera) located behind the FLiBe tank. The system would involve multiple cameras to ensure full coverage of the divertor.*

These hotspots are constructed "indentations" which provide a view of the rear of plasma facing first wall through the FLiBe tank. Compared to the rest of the VV, these hotspots would radiate a black body spectrum of a significantly higher temperature. The portion of the spectrum that lies in FLiBe's transmission band (up to 2 micron) [89,90] can be measured by remote imaging systems which look through the FLiBe from shielded locations outside the TF volume making them significantly more accessible for maintenance purposes. This could be aggregated to form a thermal image of the divertor which could be used to infer the heat flux distribution in the divertor and provide information of possible off-normal events.

## 6.4 Cherenkov radiation in FLiBe as a measure of fusion reaction rate

ARC will likely use scintillators or fission chambers located outside the FLiBe tank to detect neutrons and infer the fusion reaction rate, similar to the techniques employed today by research tokamaks. However, the possibility exists that the fusion reaction rate may also be inferred from Cherenkov radiation created in the blanket. Such a diagnostic, used in conjunction with standard techniques, may provide two advantages: (1) measurement redundancy and (2) ease of calibration. For example, plans for ITER's neutron diagnostics must consider that the intensity of neutron calibration sources are orders of magnitude less than the levels anticipated during DT plasma operation. Since detectors must be located behind the blanket shield, they may not have enough sensitivity to obtain an accurate calibration [91,92]. With the ability to detect neutrons deep or shallow in the FLiBe blanket, a Cherenkov imaging diagnostic may obtain a large dynamic range and thus a higher sensitivity to calibration sources.

The physics behind such a Cherenkov diagnostic can be described as follows: as neutrons travel through the FLiBe, they undergo inelastic nuclear reactions, which in turn lead to gamma emissions. These gammas will Compton scatter off free electrons, producing fast betas, which if moving faster than the local light speed ($2.3 \times 10^8$ m/s in FLiBe) will produce Cherenkov radiation [93]. Cherenkov radiation peaks at short wavelengths (e.g. the blue glow in fission pools), where FLiBe is a gray medium [89,90]. The total amount of Cherenkov radiation is proportional to the source neutron rate. Measurements of the Cherenkov radiation should therefore be able to closely track the time evolution of the total fusion reaction rate. An ability to monitor neutron intensity and thus neutron transport at various locations in the blanket could also be quite valuable.

Precise predictions of Cherenkov light levels are not possible at this time due to lack of data on FLiBe optical properties. However, electron energy spectra outputted from MCNP simulations indicate non-negligible fraction of electrons are above the local light speed and would provide a significant Cherenkov signal, suggesting that this would be a viable diagnostic. Figure 20 shows the electron energy spectrum (plotted in terms of velocity) normalized by the local neutron flux from MCNP simulations. The computed neutron flux per source neutron along the midplane is shown in Figure 21.

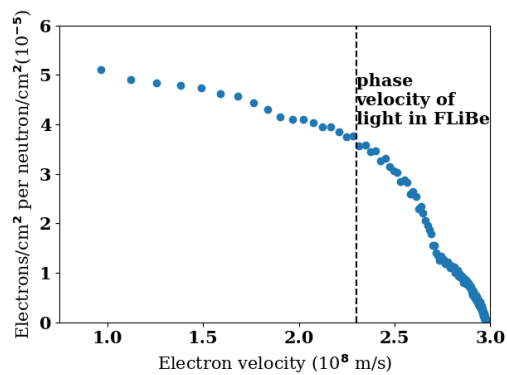

*Figure 20: Electron flux per neutron as a function of electron velocity, which is independent of position. Cherenkov radiation will be produced by electrons moving faster than the phase velocity of light in FLiBe, $2.3 \times 10^8$ m/s.*

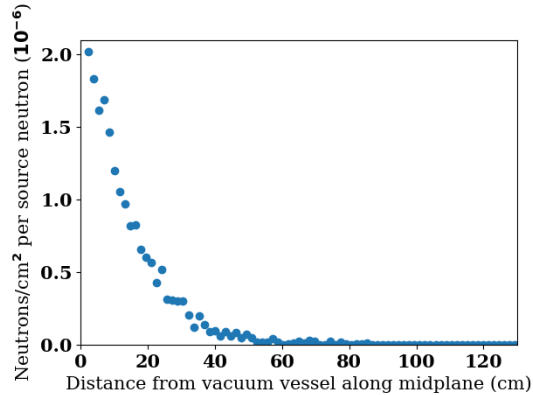

*Figure 21: Neutron flux per source neutron along the midplane in the FLiBe tank. Note that this is greater than 1 near the vessel wall because of embedded neutron multipliers in the wall.*

# 7. Future Research Needs

This conceptual design study has identified a potentially robust, integrated power exhaust management solution for the ARC pilot plant, taking advantage of the latest technological developments (e.g. HTS) and projections for advanced divertor performance (e.g. XPT divertor). However, assumptions about component performance (e.g. FLiBe immersion blanket) and the prospects of further technological developments (e.g. additive manufacturing) were used. Consequently, the realization of this vision is contingent on results from further studies and on the successful development of key technologies. Below is a summary of high-priority research identified.

*Actual performance limits of X-point target divertor*

The divertor solution implemented for ARC relies solely on model-based performance projections for long-legged divertor configurations, in particular the X-point target divertor concept. The use of these configurations therefore carries a substantial risk; these schemes have not been tested by experimentally subjecting them to reactor relevant levels of heat flux densities and plasma pressures. The operational window for a passively stable, fully detached divertor must be demonstrated to be adequate. In addition, these divertors must be shown to be compatible with excellent core plasma performance (e.g. access to high confinement regimes). Although MAST-U [16] will provide a first look at some of these issues, a divertor test tokamak, such as the proposed ADX facility [28], is required to retire these risks.

*Advanced materials, advanced manufacturing, additive manufacturing*

Construction of an ARC vacuum vessel/divertor structure that includes integrated coolant pathways and trim coils, requires the development of advanced manufacturing techniques capable of assembling large-scale, complex, multi-material-layered geometries. This will likely be enabled by additive manufacturing approaches combined with conventional fabrication and joining techniques. The design will need to account for maintenance at room temperature as well as, operation at ~800-1000 K and thus significant thermal expansion of the entire system. Advances in additive manufacturing are clearly required to implement the envisioned coolant channels, particularly those in the tungsten first wall. This development would also enable consideration of alternative coolant channel designs for further optimization. For example, optimized metal foams may provide improvements in heat transfer and structural properties compared to swirl-tubes or fins [94,95]. The development of high-performance, low activation graded materials (with appropriately high thermal conductivity and mechanical strength where needed) is also necessary for the first wall design – both for the vacuum vessel and the divertor first wall – accommodating high thermal gradients and differential thermal expansion across material interfaces.

*Properties of FLiBe*

Many open questions remain about the properties of FLiBe and resultant impact on the design. As identified in the original ARC design [1], flow-assisted erosion and corrosion of structural materials in contact with FLiBe is an area of concern, which can be further exacerbated by synergistic radiation-induced effects and influenced by the presence of a magnetic field. Erosion and corrosion properties are essential for specifying the minimum material thicknesses and for placing constraints on coolant flow, especially through coolant channels. This calls for improved data and for computational fluid dynamic models to assess the integrated cooling channel design. These would seek to optimize the competing demands for heat transfer and erosion – higher flow rates increase heat transfer rates but also increase erosion rates. In any case, a full plant model of the flow of FLiBe through the vacuum vessel and FLiBe tank is needed to more accurately assess temperature profiles and required pumping power.

In general, the lack of wavelength resolved, optical transmission and absorption data for molten FLiBe makes it difficult to assess the viability of the proposed Cherenkov and structural thermal radiation monitoring systems identified by this study. Furthermore, erosion of structural material discussed above may lead to a degradation of the transparency of FLiBe, if left untreated. Clearly, the ARC reactor must have a FLiBe conditioning system as part of its FLiBe coolant loop. Its function would be multifold: continuously remove tritium produced in the coolant; control FLiBe chemistry to minimize corrosion, and remove eroded materials. An additional unknown is the impact of neutron and gamma radiation on both the electrical and optical properties of FLiBe. MHD effects, enhanced by elevated electrical conductivity, can reduce heat transfer rates in FLiBe and adversely impact heat removal. Future research in support of the FLiBe immersion blanket concept should be targeted to address all these issues. Small scale experiments performed at temperature could be designed to simulate the FLiBe environment anticipated for ARC, but without the neutrons: testing optical properties and potential optical diagnostics; addressing corrosion effects, and FLiBe chemistry; and assessing the means to remove tritium (perhaps using deuterium as a proxy). Separate dedicated experiments could be performed to look at FLiBe's response to neutron and gamma radiation.

*Neutron flux exposure limits for HTS*

There is a clear and compelling need to test the performance of HTS in a relevant neutron flux environment and over the range of operating temperatures anticipated. Coil lifetime estimates used in this scoping study could be greatly affected with important implications for the commercial viability of this approach.

*Plasma stability and control requirements; internal trim coil design*

A development of self-consistent plasma stability and control requirements for the modified ARC design was not part of this study. As such, only crude specifications for the trim coil design and their placement could be made. Next step research should determine these requirements with a time-dependent tokamak equilibrium simulation code, assessing the need for passive stabilizers, determining optimal locations for the vertical stability trim coils and specifying current levels and time responses required. Should the use of passive stabilizers be found necessary, their impact on neutron budget, TBR and heat removal would need to be assessed.

Substantial work is needed on the internal trim coil design itself. Considerations include: AC operation and skin current effects; single-turn versus multiple turn designs; devising suitable structural supports to connect the trim coils to the vacuum vessel; schemes for attaining sufficient electrical isolation – perhaps allowing for substantial bypass current through finite-resistance 'insulators'; and developing a means to make electrical connections to outside the FLiBe tank.

*Refinements to neutron transport model; assessment of neutron shielding materials*

At present, the vacuum vessel in the MCNP model has a continuous internal structure with layers and dimensions intended for the main chamber, similar to that which was proposed in the original design study [1]. The divertor region has since been redesigned to handle the higher expected heat flux but this was not included in the MCNP model, for example the Be layer was removed in the divertor but is still present in the MCNP model. A follow-on study should explore the effect of including divertor region details, particularly as more comprehensive designs are proposed and developed. An area that needs further exploration is the time evolution of the isotopic make-up of FLiBe compared to startup. Issues include impact on TBR and a potential shift of power loading. These results could inform requirements for monitoring and replacement of FLiBe as well.

Finally, a scoping study of potential neutron shield materials for the PF coils should be performed. The present design considers 25 cm thick plates of zirconium hydride. However, concerns about chemical compatibility and safety may be alleviated with an alternative material.

*Start-up and off-normal scenarios*

There have been little to no analysis of start-up or off-normal heat flux loading scenarios, for example a shift in the magnetic equilibrium resulting in a 'limited plasma', in both design studies of ARC. These are situations that would result in significant heat flux loading on the main chamber and could potentially result in damage to the plasma facing components leading to a disruption. These transient events will affect the design of ARC and should be considered.

*Maintenance schemes and VV lifetime expectancy*

In order to fully assess the advantages of the vertical maintenance scheme and the required lifetime expectancy of the VV, a detailed study of expected downtime and cost associated with the replacement of the VV is required. Furthermore, the lifetime expectancy of the VV due to cyclic fatigue and neutron degradation needs to be assessed. This would help determine the economic viability of the ARC design.

## 8. Summary

A follow-on conceptual design study was undertaken to explore pathways for managing steady state heat exhaust from the ARC fusion pilot plant [1], which is designed to generates a fusion power output of 525 MW and total thermal output of 628 MW in a compact size ($R_0$ = 3.3 m). The demountable TF magnets enabled placing the PF coils inside the TF magnets, producing a double-null plasma equilibrium that includes a long-leg X-point target divertor geometry in both the upper and lower divertor chambers. The vacuum vessel was modified to accommodate the divertor legs without any loss to core plasma volume or increase in TF magnetic size. Augmented by neutron shielding material placed at strategic locations, the molten salt FLiBe blanket is found to adequately shield all superconducting PF coils and the demountable TF magnet, attaining the targeted lifetime of greater than 10 full power years of operation. The FLiBe blanket serves as effective medium for neutron heat removal as well as an efficient tritium breeder; advanced neutronics calculations indicate a TBR of ~1.08 for this design. The successful integration of a long-leg, X-point target divertor geometry into the ARC pilot plant design – with no impact on core plasma volume and acceptable neutron shielding and TBR – is a potentially transformative development, a significant milestone for tokamak fusion power reactor design.

The long outer divertor legs have a large surface area to accept divertor heat loads – comparable to the main chamber surface area – without making the device larger or significantly impacting tritium breeding. This feature, in conjunction with expectations that long-legged divertors can spread heat exhaust onto their sidewalls, is the key to handling the narrow heat exhaust channels that are presently projected for tokamak reactors. In addition, the extended leg geometry allows for partial neutron shielding of the divertor walls, softening the neutron spectrum and greatly reducing DPA and neutron-induced helium production rates in the divertor first wall materials. This means that existing fission reactor facilities can be used to perform relevant tests of high-heat flux component designs for the divertor region.

A tungsten swirl-tube cooling channel design is implemented in the divertor and found capable of exhausting 12 MW/m$^2$ surface heat flux, despite the high inlet temperatures of the FLiBe coolant (800K). An integrated coolant loop system is developed to deliver forced FLiBe flow (~ 2 m/s) into all vacuum vessel and divertor coolant channels and exiting into the FLiBe tank. Pumping and flow control can be performed by machinery that is external to the tank and the magnetic field. The total pumping power require to circulate FLiBe in the tank, vacuum vessel and divertor region is estimated to be ~ 3 MW, well below the targeted value of 1% of total fusion power output – a necessary requirement for an economical fusion reactor.

Finally, three novel, neutron-tolerant diagnostics were explored that take advantage of ARC's unique design features: (1) microwave interferometry to measure the detachment front locations in the divertor legs, which can be used for feedback control of vertical plasma position and upper-lower magnetic flux balance; (2) an optical IR diagnostic that look 'through' the FLiBe blanket to monitor "hotspots" on the divertor chamber walls; and (3) the monitoring of Cherenkov radiation produced in the FLiBe blanket as a means to deduce fusion power output of the reactor.

While we consider this conceptual design study to be highly successful – formulating a new vision for a potentially robust, integrated power exhaust management solution for the ARC fusion pilot plant – many questions and challenges remain. These will require further in-depth studies, innovative solutions and the continued development of cutting-edge technologies on a number of fronts – and most importantly, the further engagement of talent from the fusion research community.

## Acknowledgements

We thank Rui Viera, Leigh Ann Kesler, Lihua Zhou, Michael Wigram, Caroline Sorensen, Melanie Tetreault-Friend, Jayson Vavrek, Seung Gyou Baek, Vinny Fry, Bruce Lipschultz, and Zach Hartwig for conversations and comments that have helped to greatly improve this paper, as well as Paul Bonoli for assisting the implementation of the ACCOME code. In addition, we would like to thank the rest of the Plasma Science and Fusion Center for their assistance and advice throughout this process. This work originated from an MIT Nuclear Science and Engineering graduate course and would not have been possible without the support of the NSE Department. CAD acknowledges support from the DOE NNSA Stewardship Science Graduate Fellowship under cooperative agreement No. DE-NA0002135. EAT acknowledges support from the National Science Foundation Graduate Research Fellowship under Grant No. DGE-1122374. DGW acknowledges support of Mitsubishi Electric Research Laboratories.